\documentclass[aps,prb,reprint,twocolumn,superscriptaddress,floatfix]{revtex4-2}

\usepackage{bm}
\usepackage{amsmath,amssymb,mathtools}
\usepackage{graphicx}
\graphicspath{{./}{Figures/}}
\usepackage{multirow}
\usepackage{hyperref}
\usepackage{xcolor}
\definecolor{revisionblue}{RGB}{0,59,166}
\DeclareRobustCommand{\rev}[1]{#1}
\usepackage{subcaption}
\usepackage{esint}
\usepackage{threeparttable}
\usepackage{booktabs}
\usepackage{placeins}
\usepackage{braket}

\newcommand{\ii}{\mathrm{i}}

\newcommand{\be}{\begin{equation}}
\newcommand{\ee}{\end{equation}}
\newcommand{\Id}{\mathbf{1}_2}

\newcommand{\newreferencekey}[1]{\expandafter\def\csname newref@#1\endcsname{}}
\newreferencekey{Cui2026Spin1}
\newreferencekey{RantanenEltsov2023}
\newreferencekey{DeNittisGomi2022DIII}
\newreferencekey{AmbegaokarBaratoff1963}
\newreferencekey{AmbegaokarBaratoffErratum1963}
\NewDocumentCommand{\revisionbibitem}{o m}{%
  \normalcolor
  \IfValueTF{#1}{\originalbibitem[#1]{#2}}{\originalbibitem{#2}}%
  \ifcsname newref@#2\endcsname\fi}
\AddToHook{env/thebibliography/begin}{%
  \let\originalbibitem\bibitem\let\bibitem\revisionbibitem}

\begin{document}

\title{Quaternionic superconductivity links spinful pairing, topology, and charge-\texorpdfstring{$4e$}{4e} order}

\author{Christian Tantardini}
\email{christiantantardini@ymail.com}
\affiliation{Department of Physics and Center for the Science of Materials Berlin, Humboldt-Universit\"at zu Berlin, Zum Gro{\ss}en Windkanal 2, 12489 Berlin, Germany.}
\affiliation{Department of Materials Science and NanoEngineering, Rice University, Houston, Texas 77005, United States of America.}
\affiliation{Center for Integrative Petroleum Research, King Fahd University of Petroleum and Minerals, Dhahran 31261, Saudi Arabia.}

\author{Jacopo Masotti}
\affiliation{Hylleraas center, Department of Chemistry, UiT The Arctic University of Norway, PO Box 6050 Langnes, N-9037 Troms\o, Norway}

\author{Sabri F. Elatresh} 
\affiliation{Department of Chemistry and Chemical Biology, Cornell University, Ithaca, New York 14850-14853, United States of America.}

\author{Boris Yakobson}
\affiliation{Department of Materials Science and NanoEngineering, Rice University, Houston, Texas 77005, United States of America.}

\date{\today}

\begin{abstract}
\rev{A complex quaternion represents the singlet amplitude $\psi$ and triplet vector $\bm d$ of a superconducting gap. The scalar part of its square, $\psi^2-\bm d\cdot\bm d$, is the determinant of the pairing matrix. Its magnitude gives the product of the two pairing singular values, and its phase enters the coupling to a scalar charge-$4e$ field. A finite trilinear coupling locks the relative phase when both fields and this projection are nonzero. In a weak-coupling example, pair fluctuations lower the quartet mass, but the induced quartet quartic is negative at Gaussian order. In a square-lattice Rashba model, increasing only the triplet amplitude takes the system through a nodal interval between two gapped phases. An occupied-state Wilson loop gives the class-DIII index $\nu=0$ on the weak-triplet side and $\nu=1$ on the strong-triplet side, where a strip exhibits a helical edge pair on each boundary. On an open chain of 80 rungs the quartet correlator decays more slowly than the interorbital singlet correlator over window centres $r=5$-$10$. Both slopes grow across this interval. In a symmetric SQUID, coherent two-pair transfer alone gives an $h/4e$ period when the first current harmonic vanishes.}
\end{abstract}

\maketitle

\section{Introduction}
\label{sec:intro}

Many superconducting platforms of current interest, especially noncentrosymmetric
materials, interfacial superconductors, and spin-orbit-coupled heterostructures,
combine strong spin-orbit coupling with fermionic time-reversal symmetry
\((T^2=-1)\). In systems without inversion symmetry, parity is not a good
quantum number; antisymmetric spin-orbit coupling can lift spin degeneracy and
allow the superconducting order parameter to contain both even-parity singlet
and odd-parity triplet components~\cite{GorkovRashba2001,Frigeri2004,
Smidman2017,Fischer2023}. This setting provides a natural motivation for a
matrix-valued, spin-resolved gap structure rather than a purely scalar
Ginzburg-Landau order parameter. \rev{The usual $2\times2$ spin matrix already describes mixed pairing and its Bogoliubov-de Gennes (BdG) spectrum.\cite{Nambu1960,deGennes1966,SigristUeda1991,GorkovRashba2001} Here we use its quaternion representation to follow the same scalar combination from the two gap branches to a specific quartet coupling.}

TRI systems with $T^2=-1$ have a natural Quaternionic structure: time reversal maps the occupied subspace at $\bm k$ to that at $-\bm k$ and produces Kramers pairs at time-reversal-invariant momenta. Rotations preserving a single Kramers-pair metric form ${\rm Sp}(1)\simeq{\rm SU}(2)$, which double-covers ${\rm SO}(3)$; the corresponding group for $N$ pairs is ${\rm Sp}(N)$. Over the Brillouin zone, however, this structure relates distinct momentum fibers and must be implemented through time-reversal sewing data. The obstruction to a global time-reversal-compatible frame encodes the familiar $\mathbb Z_2$ topology and is described mathematically by the Furuta-Kametani-Matsue-Minami (FKMM) invariant.\cite{Hatsugai2010,DeNittisGomi2015CMP}

On a different front, pair-density-wave (PDW) physics has highlighted \emph{higher-order} condensation where a uniform charge-$4e$ (quartet) superconducting state can appear as a vestigial order even when the uniform $2e$ order vanishes.\cite{BergFradkinKivelson2009NatPhys,Agterberg2020ARCM} Recent theory and numerics identify microscopic routes to $4e$ phases,\cite{WuWang2024npjQM,Soldini2024PRB,PhysRevB.82.134511} and device experiments now report $4e$ supercurrents in engineered Josephson structures.\cite{Ciaccia2024CommPhys} \rev{These results motivate asking which spin scalar of a mixed gap can couple to a uniform quartet field.}

\rev{A pure phase-coherent bulk quartet condensate has flux quantum $h/4e$.\cite{BergFradkinKivelson2009NatPhys,Agterberg2020ARCM} A junction can also transfer charge $4e$ through the coherent tunnelling of two Cooper pairs. Its second current harmonic oscillates at $4eV/h$ under a direct current (DC) bias $V$ and allows Shapiro steps spaced by $hf/(4e)$ under a drive at frequency $f$.\cite{Ciaccia2024CommPhys} This transport mechanism must be distinguished from a bulk quartet condensate and from the $4\pi$ Josephson effect in topological junctions.\cite{QiHughesZhang2009PRL,ZhangKaneMele2013PRL}}

\noindent
Beyond correlated-electron platforms, phonon-mediated superhydrides under pressure provide a complementary arena with near-room-temperature $T_c$, notably LaH$_{10}$ (predicted and later synthesized) \cite{Liu2017PNAS,Somayazulu2019PRL}. Its optical response indicates strong electron-phonon coupling \cite{PhysRevB.102.024501}, and charge-density-wave physics has been discussed as part of the mechanism \cite{c7w9-7tgy}. Related high-pressure superconductivity includes elemental sulfur \cite{Struzhkin1997Nature}; a broader overview is given in Ref.~\cite{Boeri2022Roadmap}, while multicomponent superconductivity in liquid metallic hydrogen is discussed in Ref.~\cite{babaev_2024}. \rev{In the pure-singlet limit, the amplitude field reduces to
$q=\psi$, and the same notation retains the time-reversal constraints,
charge-$2e$ gauge transformation and scalar GL invariants.}

\rev{Quaternion variables have been used for several aspects of paired states. Rantanen and Eltsov use a real unit quaternion to represent the orientation of the orbital triad in superfluid $^3$He-A.\cite{RantanenEltsov2023} Cui and co-workers write the complex triplet $\bm d$ vector as a pure-vector quaternion and construct its density and spin invariants.\cite{Cui2026Spin1} In the present model the scalar component of the quaternion is an independent singlet amplitude. This permits the determinant of the physical pairing matrix to vanish through singlet-triplet cancellation even when the triplet pair-spin vector is zero.}

\rev{Our contribution is to identify this determinant with the scalar projection that enters one explicit pair-quartet free energy, and to follow that choice through its phase-locking condition and its fluctuation correction. We give the component multiplicity and field normalization in the mass correction and match the coupling in a weak-coupling example. The symmetry classification, flux quantum and Josephson frequencies used below are established results.\cite{AltlandZirnbauer1997,SchnyderRyuFurusakiLudwig2008,BergFradkinKivelson2009NatPhys,Ciaccia2024CommPhys} For the Rashba lattice model, we obtain both values of the two-dimensional DIII index from the occupied states. We derive the intervening nodal interval and identify the helical edge spectrum on the strong-triplet side. The open-chain calculation complements the infinite-chain study of Soldini, Fischer and Neupert\cite{Soldini2024PRB} by quantifying the dependence of a finite-distance comparison on system size, bond dimension and the choice of reference rung.}

\section{Quaternion preliminaries and notation}\label{sec:prelim}

Superconducting order parameters are complex fields: the overall complex
phase is the physical electromagnetic $\mathrm U(1)$ phase, while relative
complex phases between singlet and triplet components distinguish, for
example, unitary from non-unitary pairing. We therefore use the
\emph{complexified quaternion algebra} (biquaternions)
\(\mathbb H_{\mathbb C}\equiv\mathbb H\otimes_{\mathbb R}\mathbb C\)
as a convenient representation space for the spinful gap, and write
\begin{align}
q(\bm k)=\psi(\bm k)\,\mathbf 1+\sum_{i=x,y,z} d_i(\bm k)\,\boldsymbol{e}_i,
\label{eq:q-packaging:def}
\end{align}
with \(\psi,d_i\in\mathbb C\). This should be understood as a compact encoding
of the usual complex singlet-triplet gap, not as the assertion that
complexified quaternions are the only possible quaternionic way to represent a
$\mathrm U(1)$ phase. The real quaternion units \(\boldsymbol e_i\) organize
the spin structure, whereas the commuting complex unit \(\ii\) carries the
ordinary superconducting phase and the relative phases of the gap components.
\rev{Here $\bm k$ is crystal momentum and
$\bm e=(\boldsymbol e_x,\boldsymbol e_y,\boldsymbol e_z)$.
We use the bilinear contraction
$\bm d\cdot\bm d=\sum_{j=x,y,z}d_j^2$ and the Hermitian squared
magnitude $|\bm d|^2=\bm d^*\cdot\bm d=\sum_j|d_j|^2$.
The operation $\mathrm{Sc}$ extracts the coefficient of the
quaternion identity.}
The quaternionic units obey
\begin{align}
\boldsymbol{e}_i\boldsymbol{e}_j
=
-\delta_{ij}
+
\epsilon_{ijk}\boldsymbol{e}_k .
\end{align}
\rev{Here $\delta_{ij}$ is the Kronecker delta and
$\epsilon_{ijk}$ is the Levi-Civita symbol, with
$\epsilon_{xyz}=+1$; repeated Cartesian indices are summed.
The commuting complex unit satisfies $\ii^2=-1$.}

We distinguish three operations:
(i) \emph{quaternion conjugation} (i.e, flip the quaternionic vector part)
\begin{equation}
\overline q=\psi\,\mathbf 1-\sum_i d_i\,\boldsymbol{e}_i,
\end{equation}
(ii) \emph{complex conjugation}
\rev{$q^\ast=\psi^\ast\,\mathbf 1+
\sum_{i=x,y,z}d_i^\ast\,\boldsymbol e_i$},
and (iii) the \emph{Hermitian (biquaternion) involution}
\begin{equation}
q^\ddagger \equiv \overline{q^\ast}=\psi^\ast\,\mathbf 1-\sum_i d_i^\ast\,\boldsymbol{e}_i.
\label{eq:ddagger-def}
\end{equation}
It is $q^\ddagger$ (not $\overline q$) that corresponds to Hermitian adjoint in the
$2\times2$ complex matrix representation below.

\rev{Let $\sigma_{x,y,z}$ be the Pauli matrices in spin space,
$\bm\sigma=(\sigma_x,\sigma_y,\sigma_z)$, and $\Id$ the spin identity.
The space $M_2(\mathbb C)$ consists of complex $2\times2$ matrices.
We use the faithful embedding
$\varrho:\mathbb H_{\mathbb C}\to M_2(\mathbb C)$}
\begin{subequations}\label{eq:Pauli-to-quaternion}
\begin{align}
\mathbf 1 &\leftrightarrow \Id,\\
\boldsymbol{e}_x \leftrightarrow -\ii\sigma_x,\qquad
\boldsymbol{e}_y &\leftrightarrow -\ii\sigma_y,\qquad
\boldsymbol{e}_z \leftrightarrow -\ii\sigma_z.
\end{align}
\end{subequations}
Then
\begin{subequations}\label{eq:rho-properties}
\begin{align}
\varrho(q)&=\psi\,\Id-\ii\,\bm d\!\cdot\!\bm\sigma, \label{eq:rho-properties:map}\\
\varrho(q^\ddagger)&=\varrho(q)^\dagger, \label{eq:rho-properties:ddagger}\\
\det \varrho(q)&=\psi^2+\bm d\!\cdot\!\bm d, \label{eq:rho-properties:det}\\
\frac12\mathrm{Tr}\big[\varrho(q)\varrho(q)^\dagger\big]&=|\psi|^2+|\bm d|^2. \label{eq:rho-properties:trace-norm}
\end{align}
\end{subequations}
\rev{The product $q\overline q$ equals
$(\psi^2+\bm d\cdot\bm d)\mathbf 1$.
For real coefficients its scalar coefficient is the squared
quaternion magnitude. For complex amplitudes, the gauge-invariant
squared amplitude is instead $|q|_H^2=|\psi|^2+|\bm d|^2$.
Equation~\eqref{eq:rho-properties:trace-norm} expresses this quantity
as one half of the squared Hilbert-Schmidt norm of $\varrho(q)$.}

It is useful to stress the notational distinction between the commuting
complex unit \(\ii\) and the quaternion basis elements \(\boldsymbol e_i\).
The matrix \(J=\ii\sigma_y=-\varrho(\boldsymbol e_y)\) serves as the spin-\(\tfrac12\) antisymmetric metric used to contract two spinors into an \({\rm SU}(2)\)-invariant singlet.
Its role in the pairing matrix is distinct from the use of \(\boldsymbol e_y\) as a basis element of the quaternion package.

\rev{For a unit quaternion $u\in{\rm Sp}(1)$, let
$U=\varrho(u)\in{\rm SU}(2)$.
Its active spin rotation $R(U)\in{\rm SO}(3)$ is defined by
\[
U(\bm v\cdot\bm\sigma)U^\dagger
=(R(U)\bm v)\cdot\bm\sigma
\]
for every real vector $\bm v$.
The antisymmetric spinor metric $J=\ii\sigma_y$ is invariant.
The fundamental representation is pseudoreal and satisfies}
\begin{equation}
U^{\mathsf T}
=
J^{-1}U^\dagger J
=
-(\ii\sigma_y)\,U^\dagger\,(\ii\sigma_y)
=
\sigma_y U^\dagger\sigma_y .
\label{eq:SU2-transpose-symplectic}
\end{equation}
\rev{Since $J^{-1}=-J$ and $U^\dagger=U^{-1}$,
Eq.~\eqref{eq:SU2-transpose-symplectic} gives}
\begin{equation}
\rev{UJU^{\mathsf T}=J,\qquad J=\ii\sigma_y.}
\label{eq:spinor-metric-invariance}
\end{equation}
Thus the transpose in the spinor contraction is not an independent
quaternionic operation; it is fixed by the symplectic structure of the
\({\rm SU}(2)\simeq{\rm Sp}(1)\) representation.
In quaternion variables this becomes \emph{conjugation} by \(u\):
\begin{align}
q\ \longmapsto\ u\,q\,u^{-1}=\psi\,\mathbf{1}+\big(R\bm d\big)\!\cdot\!(\boldsymbol{e}_x,\boldsymbol{e}_y,\boldsymbol{e}_z),
\label{eq:spin-conjugation}
\end{align}
so the scalar part $\psi$ (singlet) is a spin scalar and the triplet $\bm d$ rotates as a vector.%
\footnote{\rev{Independent rotations of the two spin indices give
$\Delta\mapsto U_{\rm L}\Delta U_{\rm R}^{\mathsf T}$.
The matrix-gap representative defined below therefore transforms as
$q_\Delta\mapsto u_{\rm L}q_\Delta u_{\rm R}^{-1}$, where
$U_{\rm L,R}=\varrho(u_{\rm L,R})$.
A common physical spin rotation has $u_{\rm L}=u_{\rm R}$.}}

\rev{For spin-$\tfrac12$ electrons, time reversal is implemented by
$\Theta=\ii\sigma_y\mathcal K$, where $\mathcal K$
denotes complex conjugation in the chosen spin basis and
$\Theta^2=-1$.}\cite{SakuraiQM3}
\rev{In our quaternion embedding, $\ii\sigma_y=-\varrho(\boldsymbol e_y)$.
Time reversal nevertheless acts on both spin indices of the
physical pairing matrix according to
$\Delta(\bm k)\mapsto
(\ii\sigma_y)\Delta^*(-\bm k)(\ii\sigma_y)^{\mathsf T}$.
The transformation below is the resulting action on the
singlet and triplet amplitudes, not left multiplication of
$q$ by $-\boldsymbol e_y$:}
\begin{align}
q(\bm k)&=\psi(\bm k)\mathbf 1+\bm d(\bm k)\cdot\bm e\nonumber\\
&\xrightarrow{\Theta}
\psi^*(-\bm k)\mathbf 1-\bm d^*(-\bm k)\cdot\bm e\nonumber\\
&=q^\ddagger(-\bm k).
\label{eq:TR-on-q}
\end{align}
\rev{Equation~\eqref{eq:TR-on-q} gives time reversal of the
physical singlet and triplet amplitudes.
For the matrix-gap representative introduced below,
$q_\Delta=\psi\mathbf 1+\ii\,\bm d\cdot\bm e$,
the same transformation reads
$q_\Delta(\bm k)\mapsto q_\Delta^*(-\bm k)$.
Complex conjugation of the explicit factor $\ii$ in $q_\Delta$
supplies the minus sign on the physical triplet amplitude,
in agreement with Eq.~\eqref{eq:TR-on-q}.}

In the conventional notation, the spinful gap reads
\begin{align}
\Delta(\bm k)=\big[\psi(\bm k)\,\mathbf{1}_2+\bm d(\bm k)\!\cdot\!\bm\sigma\big]\,(\ii\sigma_y),
\label{eq:SU-gap}
\end{align}
where $\psi$ is the spin singlet and $\bm d$ the spin triplet.\cite{SigristUeda1991}
\rev{For this amplitude field we use the Hermitian norm}
\begin{equation}
|q(\bm k)|^2=|\psi(\bm k)|^2+|\bm d(\bm k)|^2.
\label{eq:q-packaging:norm}
\end{equation}
With the representation \eqref{eq:Pauli-to-quaternion}, the quaternion package
and the physical matrix-gap representative must be distinguished. Here
\(\mathbf 1\) denotes the quaternion identity, whereas
\(\Id\) denotes the identity matrix in spin space:
\begin{subequations}\label{eq:q-backtranslate}
\begin{align}
q_\Delta(\bm k)
&\equiv
\psi(\bm k)\,\mathbf 1+\ii\sum_{i=x,y,z}d_i(\bm k)\,\boldsymbol e_i,
\label{eq:q-backtranslate:qDelta}\\
\varrho\big(q_\Delta(\bm k)\big)
&=\psi(\bm k)\,\Id+\bm d(\bm k)\!\cdot\!\bm\sigma,
\label{eq:q-backtranslate:rho-qDelta}\\
\Delta(\bm k)
&=\varrho\big(q_\Delta(\bm k)\big)(\ii\sigma_y).
\label{eq:q-backtranslate:Delta}
\end{align}
\end{subequations}
The field \(q\) is the \({\rm Sp}(1)\)-covariant package of the singlet and
triplet amplitudes, whereas \(q_\Delta\) is the corresponding matrix-gap
representative under the embedding \(\varrho\). The extra commuting factor
\(\ii\) in \(q_\Delta\) is not an additional physical phase; it converts the
quaternion basis \(\boldsymbol e_i\mapsto-\ii\sigma_i\) into the Hermitian
Pauli basis used in the physical pairing block. 
Since \(\ii\) commutes with the quaternion units, \(\psi\) is invariant under
spin rotations, whereas the triplet \(d\)-vector transforms as the vector part
of the quaternion, i.e. as an \({\rm SO}(3)\) vector induced by
\({\rm Sp}(1)\) conjugation.

The Hermitian norm, adjoint, and spin-conjugation rules above fix the conventions used in the BdG and GL constructions below. Their relation to the global Quaternionic bundle structure is developed in Sec.~\ref{sec:topology}.\cite{Hatsugai2010,DeNittisGomi2015CMP}

\section{Quaternionic BdG: derivation and basic consequences}\label{sec:BdG}

We begin with the standard Nambu formalism, also known as the  BdG mean-field form
\begin{subequations}\label{eq:bdg-standard}
\begin{align}
\mathcal H
&=\tfrac12\sum_{\bm k}\Psi_{\bm k}^\dagger
\begin{pmatrix}
h(\bm k) & \Delta(\bm k)\\
\Delta^\dagger(\bm k) & -h^\top(-\bm k)
\end{pmatrix}
\Psi_{\bm k}, \\
\Psi_{\bm k}&=\big(c_{\bm k\uparrow},c_{\bm k\downarrow},c^\dagger_{-\bm k\uparrow},c^\dagger_{-\bm k\downarrow}\big)^{\!\top}.
\label{eq:bdg-standard:ham}
\end{align}
\end{subequations}
\rev{Here $c_{\bm k\sigma}$ annihilates an electron with
momentum $\bm k$ and spin $\sigma\in\{\uparrow,\downarrow\}$;
$\Delta$ is the gap in Eq.~\eqref{eq:SU-gap}.}
\rev{In the minimal spin-independent normal state,
$h(\bm k)=\xi_{\bm k}\Id$, where
$\xi_{\bm k}=\varepsilon_{\bm k}-\mu=\xi_{-\bm k}$
is the band energy measured from the chemical potential $\mu$.}
\cite{deGennes1966,SigristUeda1991}

\rev{Using the matrix-gap representative $q_\Delta$ defined in
Eq.~\eqref{eq:q-backtranslate:qDelta}, we adopt the
Kramers-paired Nambu spinor}
{
\begin{align}
\Phi_{\bm k}
&=W\Psi_{\bm k},
\qquad W=\operatorname{diag}(\Id,\ii\sigma_y),
\label{eq:kramers-nambu}
\end{align}
}
which rotates the hole block by $\ii\sigma_y$. In this basis, the off-diagonal pairing becomes
$\Delta'(\bm k)=\Delta(\bm k)\,(-\ii\sigma_y)=\psi\,\mathbf{1}_2+\bm d\!\cdot\!\bm\sigma$,
and the BdG matrix reads
\begin{align}
H'_{\rm BdG}(\bm k)
=\begin{pmatrix}
\xi_{\bm k}\,\mathbf{1}_2 & \ \psi\,\mathbf{1}_2+\bm d\!\cdot\!\bm\sigma\\[2pt]
\ \psi^\ast\,\mathbf{1}_2+\bm d^{\,\ast}\!\cdot\!\bm\sigma & -\xi_{\bm k}\,\mathbf{1}_2
\end{pmatrix}.
\label{eq:bdg-rotated}
\end{align}

\rev{Let $\tau_{x,y,z}$ be the Pauli matrices in Nambu space,
$\tau_0$ its identity, and $\tau_\pm=(\tau_x\pm\ii\tau_y)/2$.
In this basis we write $H_{\rm BdG}\equiv H'_{\rm BdG}$,
with the prime used when emphasizing the basis rotation.
Equation~\eqref{eq:bdg-rotated} becomes}
\begin{align}
H_{\rm BdG}(\bm k)
=
\xi_{\bm k}\,\tau_z
+\tau_+\,q_\Delta(\bm k)
+\tau_-\,q_\Delta^\ddagger(\bm k),
\label{eq:BdGq-main}
\end{align}
where \(q_\Delta\) and \(q_\Delta^\ddagger\) act on spin via
\(\varrho(\cdot)\) [Sec.~\ref{sec:prelim}]. This compresses the physical
spinful pairing block
\(\Delta'=\psi\Id+\bm d\cdot\bm\sigma\)
into a single quaternionic representative. \rev{The eigenvalues of $\Delta'\Delta'^\dagger$ are
the squared pairing singular values entering the
BdG spectrum.}

\rev{For real $\psi$ and $\bm d$, the eigenvalues of $\Delta'$
are $\psi\pm|\bm d|$.
The two pairing singular values differ wherever $\psi|\bm d|\ne0$.
For general complex amplitudes,}
\begin{align}
\Delta'(\bm k)\Delta'^\dagger(\bm k)
=\Big(|\psi|^2+|\bm d|^2\Big)\Id
+\bm m(\bm k)\!\cdot\!\bm\sigma,
\label{eq:DDdagger-decomp}
\end{align}
with
\begin{equation}
\bm m(\bm k)=\psi\,\bm d^\ast+\psi^\ast\,\bm d+\rev{\ii}\,\bm d\times\bm d^\ast.
\label{eq:m-vector}
\end{equation}
Therefore,
\begin{equation}
\begin{aligned}
E_{\bm k,s,b}
&=s\sqrt{\xi_{\bm k}^2+\lambda_b(\bm k)},\\
\lambda_b(\bm k)
&=|\psi|^2+|\bm d|^2+b|\bm m|,
\qquad s,b\in\{+1,-1\}.
\end{aligned}
\label{eq:spec}
\end{equation}
Here \(s\) labels the particle-hole sign and \(b\) labels
the two gap branches; \(\lambda_\pm\equiv\lambda_{b=\pm1}\).
\rev{Here \(q=\psi\mathbf 1+\bm d\cdot\bm e\) denotes the
singlet-triplet amplitude field. The rotated pairing block is
\(\Delta'=\varrho(q_\Delta)\), with
\(q_\Delta=\psi\mathbf 1+\ii\,\bm d\cdot\bm e\).
Its determinant equals the scalar part of \(q^2\).}
{
\begin{equation}
\begin{aligned}
\det\Delta' &= \psi^2-\bm d\!\cdot\!\bm d=\mathrm{Sc}(q^2),\\
\lambda_+\lambda_- &= \det(\Delta'\Delta'^\dagger)
                    = |\mathrm{Sc}(q^2)|^2 .
\end{aligned}
\label{eq:gap-determinant}
\end{equation}}
\rev{Thus the magnitude of the scalar selected for quarteting is the product
of the two singular values of the physical pairing matrix. At a momentum
where \(q\ne0\), its vanishing means that one pairing singular value is
zero. The excitation energy in the spin-independent model also involves
\(\xi_{\bm k}\), as in Eq.~\eqref{eq:spec}.}
The simpler ``perfect-square'' spectrum
\begin{equation}
E_{\bm k}=\pm\sqrt{\xi_{\bm k}^2+|\psi|^2+|\bm d|^2}
\label{eq:spec-unitary}
\end{equation}
holds if and only if the state is \emph{unitary},
\begin{equation}
\bm m(\bm k)=0
\quad\Longleftrightarrow\quad
\Delta'(\bm k)\Delta'^\dagger(\bm k)\propto \Id .
\label{eq:unitary-condition}
\end{equation}
In the spin-independent normal-state model, unitarity gives equal gap magnitudes for the two branches at fixed momentum. Equivalently, the quaternion product $q_\Delta q_\Delta^\ddagger$ has only a scalar part. When $\bm m\ne0$, the branches split according to Eq.~\eqref{eq:spec}. For pure triplet pairing, $\bm m=\ii\bm d\times\bm d^*$ measures pair-spin polarization; for mixed pairing it also contains singlet-triplet interference, so a nonzero value at fixed momentum need not imply net polarization or broken time reversal.

Replacing the physical triplet coefficient by $\ii\bm d$ with real $\psi$ and $\bm d$ would produce a unitary pairing block, but changes the relative physical phase. It is not the algebraic conversion between $q$ and $q_\Delta$ and need not satisfy the time-reversal constraints of Eqs.~\eqref{eq:TRS-Delta} and \eqref{eq:TRS-psi-d}.

\rev{For $b=\pm1$, let
$E_b(\bm k)=\sqrt{\xi_{\bm k}^2+\lambda_b(\bm k)}>0$.
In singular channel $b$, the particle and hole weights of the
positive-energy BdG eigenstate are}
{
\begin{equation}
\begin{aligned}
|u_{\bm k,b}|^2
&=\frac12\left(1+\frac{\xi_{\bm k}}{E_b(\bm k)}\right),\\
|v_{\bm k,b}|^2
&=\frac12\left(1-\frac{\xi_{\bm k}}{E_b(\bm k)}\right).
\end{aligned}
\label{eq:uv}
\end{equation}
}

\section{Symmetries and the ten-fold way in quaternion variables}
\label{sec:tenfold}

\rev{We first establish the symmetries in the complex matrix
representation and then translate them into quaternion variables.
The matrices $\tau_i$ act in Nambu space, whereas $\sigma_i$
act in spin space.}

\rev{For a spin-dependent normal state, the electron Hamiltonian is}
\begin{equation}
h(\bm k)
=
\xi_{\bm k}\Id
+
\bm g(\bm k)\cdot\bm\sigma
+
\bm B\cdot\bm\sigma .
\label{eq:normal-h-spin}
\end{equation}
\rev{Here $\bm g(\bm k)$ is the spin-orbit splitting vector,
and $\bm B$ is the Zeeman energy vector, including the conversion
from magnetic field to energy. In the GL expressions below,
$\bm B$ instead denotes the physical magnetic field, with this
conversion included in the phenomenological couplings.}

\rev{The basis rotation
$W=\mathrm{diag}(\Id,\ii\sigma_y)$ gives the Kramers-Nambu
Hamiltonian}
\begin{equation}
H'_{\rm BdG}(\bm k)
=
\begin{pmatrix}
h(\bm k) & \Delta'(\bm k)\\
\Delta'^\dagger(\bm k)
&-\sigma_y h^{\mathsf T}(-\bm k)\sigma_y
\end{pmatrix},
\label{eq:BdG-normal-gap-explicit}
\end{equation}
\rev{where $\Delta'=\psi\Id+\bm d\cdot\bm\sigma$ and
$\Delta=\Delta'\ii\sigma_y$ is the physical pairing matrix.}

\rev{The intrinsic particle-hole operator must undergo the same
basis transformation. In the conventional basis,
$\mathcal C_0=(\tau_x\otimes\Id)\mathcal K$.
An antiunitary operator $U\mathcal K$ transforms as
$WUW^{\mathsf T}\mathcal K$, giving}
\rev{
\begin{align}
\mathcal C H'_{\rm BdG}(\bm k)\mathcal C^{-1}
&=-H'_{\rm BdG}(-\bm k),\nonumber\\
\mathcal C&=(\tau_y\otimes\sigma_y)\mathcal K,
\qquad \mathcal C^2=+1.
\label{eq:phs}
\end{align}
}
\rev{Here $\mathcal K$ complex-conjugates the matrix entries
in the chosen spin and Nambu basis at fixed momentum;
momentum reversal appears explicitly in the symmetry relation.
Hermiticity fixes the lower off-diagonal block to
$\Delta'^\dagger$. Particle-hole symmetry additionally requires
fermionic antisymmetry of the gap and the corresponding
hole-sector normal Hamiltonian.}

\rev{Physical time reversal is represented in this basis by
$\mathcal T=(\tau_0\otimes\ii\sigma_y)\mathcal K$, with
$\mathcal T^2=-1$. It is a symmetry only when both the normal
and pairing sectors satisfy their time-reversal constraints.
The normal-state condition is
$h(-\bm k)=\sigma_y h(\bm k)^*\sigma_y$.
For the pairing block, fermionic antisymmetry and time reversal
give, respectively,}
\begin{subequations}\label{eq:AZ-on-q}
\begin{align}
\Delta'(\bm k)
&=\sigma_y\Delta'^{\mathsf T}(-\bm k)\sigma_y,
\label{eq:PHS-Delta}\\
\Delta'(\bm k)
&=\sigma_y\Delta'^*(-\bm k)\sigma_y.
\label{eq:TRS-Delta}
\end{align}
\end{subequations}
\rev{The first relation follows from
$\Delta(\bm k)=-\Delta^{\mathsf T}(-\bm k)$.
The second is imposed in a time-reversal-invariant gauge.
Thus transpose, complex conjugation and Hermitian adjoint
enter for different reasons: Fermi antisymmetry, antiunitarity
and Hermiticity.}

\rev{We now replace the spin matrices by quaternion entries,
while retaining the two-dimensional Nambu block structure.
The embedding $\varrho$ is extended entrywise to these block
matrices. For any biquaternion $a$, our convention
$\varrho(\boldsymbol e_i)=-\ii\sigma_i$ gives}
\rev{
\begin{align*}
\sigma_y\varrho(a)^{\mathsf T}\sigma_y
&=\varrho(\overline a),\\
\sigma_y\varrho(a)^*\sigma_y
&=\varrho(a^*).
\end{align*}
}
\rev{Here $\overline a$ reverses the quaternion vector part,
whereas $a^*$ conjugates its complex coefficients.
These identities provide the translation of the matrix
constraints into quaternion operations.}

\rev{Define the normal-state representative
$h_q(\bm k)=\xi_{\bm k}\mathbf 1+
\ii[\bm g(\bm k)+\bm B]\cdot\bm e$.
Together with
$q_\Delta=\psi\mathbf 1+\ii\bm d\cdot\bm e$,
it yields}
\rev{
\begin{align*}
\mathsf H(\bm k)
&=
\begin{pmatrix}
h_q(\bm k) & q_\Delta(\bm k)\\
q_\Delta^\ddagger(\bm k)
&-\overline{h_q(-\bm k)}
\end{pmatrix},
&
H'_{\rm BdG}(\bm k)&=\varrho\!\left(\mathsf H(\bm k)\right).
\end{align*}
}
\rev{The quaternion conjugation in the hole block follows
from the transpose in
Eq.~\eqref{eq:BdG-normal-gap-explicit}.
The adjoint $q_\Delta^\ddagger$ ensures Hermiticity;
no independent hole-gap field is introduced.}

\rev{Here $\mathcal C$ is the intrinsic BdG particle-hole
operator and $\mathcal T$ is physical time reversal.
When the full superconducting Hamiltonian has unbroken
spin ${\rm SU}(2)$ symmetry, composing $\mathcal C$ with
a spin rotation by $\pi$ gives the effective particle-hole
operator $\widetilde{\mathcal C}$, with
$\widetilde{\mathcal C}^{\,2}=-1$.
We include this operator to describe the spin-reduced
classes C and CI. Their quaternionic Nambu-block
representations are:}
\begin{subequations}\label{eq:AZ-ops}
\rev{
\begin{align}
\mathcal C
&=\varrho\!\begin{pmatrix}
0 & \boldsymbol e_y\\
-\boldsymbol e_y & 0
\end{pmatrix}\mathcal K,
&\mathcal C^2&=+1,
\label{eq:AZ-ops:Cplus}\\
\widetilde{\mathcal C}
&=\varrho\!\begin{pmatrix}
0 & \mathbf 1\\
-\mathbf 1 & 0
\end{pmatrix}\mathcal K,
&\widetilde{\mathcal C}^{2}&=-1,
\label{eq:AZ-ops:Cminus}\\
\mathcal T
&=\varrho\!\begin{pmatrix}
-\boldsymbol e_y & 0\\
0 &-\boldsymbol e_y
\end{pmatrix}\mathcal K,
&\mathcal T^2&=-1.
\label{eq:AZ-ops:T}
\end{align}
}
\end{subequations}
\rev{These are operators on the complex Nambu spinor:
$\mathcal K$ acts after the embedding and is not
coefficient-wise quaternion conjugation.
Since $\varrho(\boldsymbol e_y)^*=
\varrho(\boldsymbol e_y)$, all three block matrices
are real after embedding. Their antiunitary squares
therefore follow directly from ordinary block multiplication
and $\boldsymbol e_y^2=-\mathbf 1$.
The operator $\widetilde{\mathcal C}$ is a symmetry only
under the additional spin-rotation assumption discussed below.}

\rev{The pairing constraints
\eqref{eq:PHS-Delta} and \eqref{eq:TRS-Delta}
become, respectively,}
\rev{
\begin{align*}
q_\Delta(\bm k)&=\overline{q_\Delta(-\bm k)},\\
q_\Delta(\bm k)&=q_\Delta^*(-\bm k).
\end{align*}
}
\rev{Equivalently, their component forms are}
\begin{subequations}\label{eq:AZ-on-q-quat}
\begin{align}
\psi(\bm k)&=\psi(-\bm k),\qquad
\bm d(\bm k)=-\bm d(-\bm k),
\label{eq:parity-psi-d}\\
\psi(\bm k)&=\psi^*(-\bm k),\qquad
\bm d(\bm k)=-\bm d^*(-\bm k).
\label{eq:TRS-psi-d}
\end{align}
\end{subequations}
\rev{For the amplitude package
$q=\psi\mathbf 1+\bm d\cdot\bm e$, physical time reversal
therefore induces $q(\bm k)\mapsto q^\ddagger(-\bm k)$,
as in Eq.~\eqref{eq:TR-on-q}.
The different quaternion operations on $q$ and $q_\Delta$
reflect their different definitions.
The normal-state time-reversal condition becomes
$h_q(-\bm k)=h_q^*(\bm k)$.}

\rev{A common physical spin rotation acts by
${\rm Sp}(1)$ conjugation on $q$, $q_\Delta$ and $h_q$,
rotating the vectors $\bm d$, $\bm g$ and $\bm B$
together. A gauge rotation multiplies the particle and
hole Nambu components by $e^{\ii\phi}$ and
$e^{-\ii\phi}$, respectively, and sends}
\begin{align}
q(\bm k)&\mapsto e^{2\ii\phi}q(\bm k),
\qquad
q_\Delta(\bm k)\mapsto e^{2\ii\phi}q_\Delta(\bm k).
\label{eq:gauge-on-q}
\end{align}
\rev{Thus both fields carry charge $2e$ in our gauge
convention, with $e>0$ the elementary charge magnitude.
With $\hbar=1$ and $\bm A$ the electromagnetic vector
potential, the pair and quartet covariant gradients are
$(\boldsymbol\nabla-2\ii e\bm A)q$ and
$(\boldsymbol\nabla-4\ii e\bm A)Q$, respectively.}

\rev{Equation~\eqref{eq:AZ-ops:Cplus} expresses the intrinsic
BdG particle-hole redundancy. If the complete superconducting
Hamiltonian has unbroken spin ${\rm SU}(2)$ symmetry, the
spin rotation}
\rev{
\[
R_y(\pi)=
\varrho\!\begin{pmatrix}
\boldsymbol e_y&0\\
0&\boldsymbol e_y
\end{pmatrix}
\]
}
\rev{is also a symmetry. Its composition with PHS gives
$\widetilde{\mathcal C}=-R_y(\pi)\mathcal C$,
where the overall minus sign is an irrelevant phase.
When physical time reversal is present, the same rotation gives
$\widetilde{\mathcal T}=R_y(\pi)\mathcal T=\mathcal K$,
with $\widetilde{\mathcal T}^{2}=+1$.
The resulting spin-reduced classes are C and CI, respectively.
This reduction is unavailable for generic spin-orbit coupling
or spin-symmetry-breaking triplet order.}

\rev{With intrinsic PHS and physical TRS, the unitary product
$\mathcal T\mathcal C$ anticommutes with the Hamiltonian and,
up to an overall phase, defines the chiral operator
$\mathcal S$. The symmetry signs of the complete normal
and pairing Hamiltonian determine the AZ class.
Generic spin-mixed time-reversal-invariant superconductors
belong to DIII.}
\cite{AltlandZirnbauer1997,SchnyderRyuFurusakiLudwig2008,
Kitaev2009AIP,RyuSchnyderFurusakiLudwig2010NJP,
ChiuTeoSchnyderRyu2016RMP}

\rev{Table~\ref{tab:AZ-combined} summarizes the corresponding
stable free-fermion classifications. Quaternionic building
blocks can also be used to construct lattice models in
specified AZ classes, as demonstrated by Deng \emph{et al.}}
\cite{DengWangDuan2014PRB}

\begin{table*}[t]
\caption{Symmetry content and topological classification of selected Altland-Zirnbauer classes.}
\label{tab:AZ-combined}
\centering
\begin{tabular*}{\textwidth}{@{\extracolsep{\fill}}lcccc@{}}
\toprule
\textbf{AZ class} &
\textbf{TRS} &
\textbf{PHS} &
\textbf{Chiral} &
\textbf{Topological group $(d=1,2,3)$} \\
\midrule
D
& $\times$
& $\mathcal C^2=+1$
& $\times$
& $(\mathbb Z_2,\mathbb Z,0)$ \\
C
& $\times$
& $\widetilde{\mathcal C}^2=-1$
& $\times$
& $(0,2\mathbb Z,0)$ \\
DIII
& $\mathcal T^2=-1$
& $\mathcal C^2=+1$
& $\checkmark$
& $(\mathbb Z_2,\mathbb Z_2,\mathbb Z)$ \\
CI
& \rev{$\widetilde{\mathcal T}=\mathcal K$,
$\widetilde{\mathcal T}^{2}=+1$}
& $\widetilde{\mathcal C}^2=-1$
& $\checkmark$
& $(0,0,2\mathbb Z)$ \\
\bottomrule
\end{tabular*}
\vspace{1ex}

\begin{minipage}{\linewidth}
\footnotesize
\textit{Note.}
\rev{Intrinsic PHS requires fermionic antisymmetry of the gap
[Eq.~\eqref{eq:PHS-Delta}] and the corresponding hole block
in Eq.~\eqref{eq:BdG-normal-gap-explicit}.
Physical TRS requires both $h_q(-\bm k)=h_q^*(\bm k)$
and the pairing condition in Eq.~\eqref{eq:TRS-psi-d},
in a time-reversal-invariant gauge.
Classes C and CI assume unbroken spin ${\rm SU}(2)$ symmetry
and the corresponding spin-reduced representation;
in class CI the effective time-reversal operator is
$\widetilde{\mathcal T}=\mathcal K$.}
\end{minipage}
\end{table*}

\section{Ginzburg-Landau theory with a quaternion order parameter}\label{sec:GL}

In the GL expressions we set $\hbar=1$; $F$ denotes a physical free energy and its thermal statistical weight is $\exp[-F/(k_B T)]$. \rev{Here $d$ is the spatial dimension and $\bm r$ the position coordinate.}
We use the Hermitian norm
\(|q|^2\equiv |q|_H^2\), defined in
\eqref{eq:rho-properties:trace-norm}; equivalently,
\(|q|_H^2=\mathrm{Sc}(q^\ddagger q)\). \rev{As a minimal model near the pair transition, we take}
\begin{align}
F[q]=\int d^d\mathbf r\,
\Big\{
\alpha\,|q|^2
+\beta_1\,|q|^4
+\kappa_1\,|(\boldsymbol D q)|^2
\Big\},
\label{eq:GLq:iso}
\end{align}
with \(\boldsymbol D=\boldsymbol\nabla-2\ii e\,\boldsymbol A\).
\rev{Equation~\eqref{eq:GLq:iso} assigns the singlet and triplet the same bare transition temperature and the same stiffness. In a spin-rotation-invariant system the two quadratic coefficients can instead be independent, \(\alpha_s(T)|\psi|^2+\alpha_t(T)|\bm d|^2\), with separate stiffness tensors \(K^s_{ij}\) and \(K^t_{ij}\). The common-coefficient choice gives the four pair components the same Gaussian propagator in Sec.~\ref{sec:4e}.}\cite{SigristUeda1991}

Beyond the isotropic \( |q|^4 \) term, we retain two additional symmetry-allowed quartic structures that compactly represent selected singlet-triplet invariants. These
quartic structures are built from two basic covariant quantities: the scalar
part of \(q^2\), which gives the invariant \(|\mathrm{Sc}(q^2)|^2\), and the
\rev{triplet pair-spin vector} \(\bm N\), which gives the invariant \(|\bm N|^2\).
\rev{The scalar projection is given in Eq.~\eqref{eq:gap-determinant}.
The triplet pair-spin vector is}
\begin{equation}
\bm N=\ii\,\bm d\times\bm d^*.
\label{eq:GLq:Nvec}
\end{equation}
\rev{The vector $\bm N$ measures non-unitarity of the triplet
sector.}
{
The component magnitudes and relative phases determine $\bm N$,
which is invariant under a common phase rotation.
At each momentum and position,
\[
\bm N=0\quad\Longleftrightarrow\quad
\bm d=e^{\ii\phi}\bm n,\qquad \bm n\in\mathbb R^3.
\]
Thus a phase choice makes the triplet vector real at that point.
A single global gauge rotation makes the entire texture real only
when this phase is common throughout the three components of $\bm d$ in the texture.
}
\rev{The full mixed gap is unitary when \(\bm m=\bm N+2\,\mathrm{Re}(\psi^*\bm d)=0\) [Eq.~\eqref{eq:unitary-condition}]. For example, at a fixed momentum take \(\psi=1\) and \(\bm d=(0,0,1)\). Then \(\bm N=0\), while \(\bm m=(0,0,2)\), \((\lambda_+,\lambda_-)=(4,0)\), and \(\mathrm{Sc}(q^2)=0\). The gap splitting here comes entirely from singlet-triplet interference. In the spin-matrix description of Sigrist and Ueda, the singular values characterize the pairing gap.\cite{SigristUeda1991} Fernandes and Fu obtain the charge-\(4e\) scalar $\Delta_1^2+\Delta_2^2$ from the two lattice components of a nematic pair field.\cite{FernandesFu2021PRL} In our singlet-triplet model, Eq.~\eqref{eq:gap-determinant} identifies the particular scalar coupled to \(Q\) with a determinant of the physical gap.}

The restricted quartic functional is then
\begin{align}
F_4[q]=\int d^d\mathbf r\,
\Big\{
\beta_1\,|q|^4
+\beta_2\,\big|\mathrm{Sc}(q^2)\big|^2
+\beta_3\,|\bm N|^2
\Big\}.
\label{eq:GLq:quartic}
\end{align}
\rev{For one spin scalar \(\psi\) and one spin vector \(\bm d\), define \(S=|\psi|^2\), \(D=\bm d^*\!\cdot\!\bm d\), and \(P=\bm d\!\cdot\!\bm d\). At zero field, spin rotations, gauge invariance, and time reversal allow five independent real homogeneous quartic invariants,}
{
\[
 S^2,\qquad D^2,\qquad |P|^2,\qquad SD,\qquad
 \mathrm{Re}(\psi^{*2}P).
\]}
\rev{The two triplet contractions are \(D^2\) and \(|P|^2\); the last term is the time-reversal-even combination that depends on the relative phase. Equation~\eqref{eq:GLq:quartic} retains three combinations, \((S+D)^2\), \(|\psi^2-P|^2\), and \(|\bm N|^2=D^2-|P|^2\). Its three coefficients therefore describe a restricted subset of the spin-isotropic theory. Further orbital components or spin-orbit coupling require invariants appropriate to the crystal representation.}\cite{SigristUeda1991}

Crystal anisotropy appears via a symmetric stiffness tensor. To avoid double-counting, combine the isotropic and anisotropic pieces into
\begin{align}
F_{\mathrm{grad}}[q]=\int d^d\mathbf r\,
\sum_{ij} K_{ij}\,
\mathrm{Sc}\!\big[(D_i q)(D_j q)^\ddagger\big].
\label{eq:GLq:grad}
\end{align}
\rev{We take $K_{ij}=\kappa_1\delta_{ij}+\kappa_{ij}$ real,
symmetric and positive definite in the gradient-stable regime
considered here. The point group constrains the anisotropic part
$\kappa_{ij}=\kappa_{ji}$.} Here
\(D_i q=(\partial_i-2\ii e A_i)q\), so
\((D_i q)^\ddagger=(\partial_i+2\ii e A_i)q^\ddagger\). In quaternion
variables, \(K_{ij}(D_i q)(D_j q)^\ddagger\) is manifestly
\({\rm Sp}(1)\)-covariant and retains a common spatial stiffness tensor; additional component-dependent gradients are allowed in a general crystal.\cite{SigristUeda1991}

In noncentrosymmetric systems with SOC, a magnetic field permits magnetoelectric invariants linear in gradients,
\begin{align}
F_{\mathrm{LI}}[q]=\int d^d\mathbf r\,
\sum_{ij}\gamma_{ij}B_i\,
\mathrm{Im}\,\mathrm{Sc}\!\left[q^\ddagger D_jq\right],
\label{eq:GLq:LI}
\end{align}
where the point group constrains $\gamma_{ij}$. The current-like bilinear and $\bm B$ are both odd under time reversal, so their product is allowed. For a Rashba axis $\hat{\bm n}$ the invariant has the form $(\hat{\bm n}\times\bm B)\cdot\mathrm{Im}\,\mathrm{Sc}(q^\ddagger\boldsymbol Dq)$ and can favor a field-induced helical phase. A fixed time-reversal-even polar vector alone cannot multiply this current-like bilinear in a time-reversal-invariant free energy.\cite{Edelstein1995PRL,Frigeri2004,KaurAgterbergSigrist2005PRL}

\rev{A magnetic field can contribute
$f_B=\eta_B\,\bm B\cdot\bm N$
to the free-energy density, where $\eta_B$ is the phenomenological
Zeeman coupling to the triplet pair spin.}
\cite{SigristUeda1991}
Gauge couplings enter only through $\boldsymbol D$ in ~\eqref{eq:GLq:iso}-\eqref{eq:GLq:grad}; all quaternion invariants are built to be ${\rm U}(1)$-covariant.

In weak coupling one recovers the standard scalings
\begin{subequations}\label{eq:GLq:weak}
\begin{align}
\alpha(T) &= a_{\rm GL}\!\left(\frac{T}{T_c}-1\right), \label{eq:GLq:weak-alpha}\\
\beta_1 &\sim \frac{7\,\zeta(3)}{8\pi^2 (k_B T_c)^2}\,N(0), \label{eq:GLq:weak-beta}\\
\kappa_1 &\sim \frac{7\,\zeta(3)}{48\pi^2 (k_B T_c)^2}\,N(0)\,\langle v_F^2\rangle, \label{eq:GLq:weak-kappa}
\end{align}
\end{subequations}
\rev{These expressions give weak-coupling scales for gap amplitudes
in energy units; their numerical prefactors depend on the order
parameter and density-of-states conventions.
Here $T_c$ is the common bare pair transition temperature,
$a_{\rm GL}=T_c(\partial\alpha/\partial T)_{T_c}>0$,
and $N(0)=N_F$ is the Fermi-level density of states per spin.
The quantity $v_F$ is the Fermi speed, and
$\langle v_F^2\rangle$ is the normalized Fermi-surface average
of its square.
The displayed gradient prefactor is for an isotropic
three-dimensional Fermi surface.
In the two-pocket example of Appendix~\ref{app:analytic},
$\beta_1=2b$ and $\beta_2=-b$, so the pure-singlet quartic
coefficient is $\beta_1+\beta_2=b$.}
\cite{TinkhamBook}
\rev{Equations~\eqref{eq:gap-determinant}, \eqref{eq:GLq:Nvec} and \eqref{eq:GLq:quartic}} thus compress singlet-triplet bookkeeping into two scalars, $|q|^2$ and $|\mathrm{Sc}(q^2)|^2$, plus the vector $\bm N$, keeping TR, ${\rm Sp}(1)$, and point-group constraints visible in the same variables. When inversion is broken, the Lifshitz piece \eqref{eq:GLq:LI} sits directly next to the usual gradient energy, and its field-induced helical response depends on the magnetoelectric tensor and stiffness.

\section{\texorpdfstring{\rev{Topological criteria for the occupied BdG states}}{Topological criteria for the occupied BdG states}}\label{sec:topology}

\rev{We use the time-reversal structure of the negative-energy BdG states to calculate the two-dimensional DIII index. In Sec.~\ref{sec:numerics}, a change in the triplet amplitude switches the index of the same lattice Hamiltonian. Appendix~\ref{secS:edge} gives an independent analytic classification.}

\rev{For a time-reversal-invariant spinful BdG Hamiltonian with
$\mathcal T^2=-1$ and a bulk gap at zero energy, the negative-energy
subspaces form a Quaternionic bundle under $\bm k\mapsto-\bm k$.} The antiunitary map between these fibers squares to $-1$. This global structure does not imply that an arbitrary local Berry connection is pointwise $\mathfrak{sp}(N)$-valued, nor that a global smooth periodic time-reversal-compatible frame exists in a nontrivial phase.\cite{DeNittisGomi2015CMP,ChiuTeoSchnyderRyu2016RMP}

\rev{Let $2N$ be the number of negative-energy bands and
$\{|u_n(\bm k)\rangle\}_{n=1}^{2N}$ a local orthonormal frame
for them. With $\partial_i\equiv\partial/\partial k_i$,
the Berry connection and curvature are}
\begin{subequations}\label{eq:quat-connection}
\begin{align}
[\mathcal A_i(\bm k)]_{mn}&=\langle u_m(\bm k)|\partial_{k_i}u_n(\bm k)\rangle,
\qquad \mathcal A_i\in\mathfrak u(2N),\\
\mathcal F_{ij}&=\partial_i\mathcal A_j-\partial_j\mathcal A_i+[\mathcal A_i,\mathcal A_j].
\end{align}
\end{subequations}
Time reversal relates the frames at opposite momenta through the sewing matrix
\begin{equation}
w_{mn}(\bm k)=\langle u_m(-\bm k)|\mathcal T|u_n(\bm k)\rangle.
\label{eq:sewing-matrix}
\end{equation}
The resulting transformation laws include $w$ and its derivatives. In particular, TRS makes the trace of the Berry curvature odd under momentum reversal, so the total first Chern number in two dimensions vanishes. The remaining strong class-DIII invariants are $\mathbb Z_2$ in two dimensions and $\mathbb Z$ in three dimensions.\cite{SchnyderRyuFurusakiLudwig2008,ChiuTeoSchnyderRyu2016RMP}

\rev{For the two-dimensional index, choose a smooth periodic complex
frame of occupied states over the Brillouin zone (BZ).
Such a complex frame exists because the total first Chern number
vanishes and need not respect time reversal.
At the four time-reversal-invariant momenta (TRIM)
$\Lambda_\alpha$, $\alpha=1,\ldots,4$, the sewing matrix is
antisymmetric. With $\mathrm{Pf}$ denoting the Pfaffian,
the strong index is}
\begin{equation}
(-1)^\nu=\prod_\alpha
\frac{\mathrm{Pf}\,w(\Lambda_\alpha)}{\sqrt{\det w(\Lambda_\alpha)}}.
\label{eq:FK-Pfaffian_BdG}
\end{equation}
The square-root branches must be fixed consistently by continuous sewing data; independently choosing a principal square root at each TRIM is insufficient. The invariant depends on the full occupied BdG eigenvectors, including the normal-state dispersion and SOC. \rev{Equivalently, the strong two-dimensional DIII index is the $\mathbb Z_2$ index of the occupied Quaternionic bundle. We evaluate it from the spectral flow of non-Abelian Wilson loops over half the Brillouin zone.\cite{FuKane2006TRPolarization,QiHughesZhang2009PRL,SoluyanovVanderbilt2011,YuQiBernevigFangDaiVanderbilt2011,DeNittisGomi2022DIII}}

For the three-dimensional winding invariant, first diagonalize the chiral operator and write the full gapped Hamiltonian as
\begin{equation}
\begin{aligned}
H_{\rm ch}(\bm k)&=\begin{pmatrix}0&D(\bm k)\\D^\dagger(\bm k)&0\end{pmatrix},\\
U&=D(D^\dagger D)^{-1/2}\in{\rm U}(n).
\end{aligned}
\label{eq:full-chiral-flattening}
\end{equation}
Here $n$ is the dimension of one chiral sector. The block $D$ contains both the normal and pairing terms. Its polar unitary $U$, rather than the normalized pairing package $q/|q|$, enters the winding number:
\begin{align}
\nu_3={}&\frac{1}{24\pi^2}\int_{\rm BZ}d^3k\,\epsilon^{ijk}\nonumber\\
&\times\mathrm{tr}\!\left[(U^{-1}\partial_iU)(U^{-1}\partial_jU)(U^{-1}\partial_kU)\right].
\label{eq:DIII-winding}
\end{align}
\rev{Repeated momentum indices are summed with $\epsilon^{xyz}=+1$.
Reversing the chiral-sector orientation changes the sign of $\nu_3$.} In suitable minimal models $U$ can be represented by a unit real quaternion in ${\rm Sp}(1)\simeq{\rm SU}(2)$; normalizing a generic complexified quaternion by its Hermitian norm does not establish this property. In the free-fermion classification, $|\nu_3|$ gives the minimal number of protected surface Majorana cones.\cite{SchnyderRyuFurusakiLudwig2008,ChiuTeoSchnyderRyu2016RMP}

In inversion-symmetric systems, parity data and the pairing parity can simplify the index under the assumptions of the relevant parity criterion. Weak-pairing formulas instead use the projected gap signs and the topology of normal-state Fermi surfaces.\cite{FuKane2007Inversion,FuBerg2010OddParity,Sato2010OddParity,QiHughesZhang2009PRL} For nodal systems, momentum-resolved invariants must be evaluated on gapped loops or slices of the full Hamiltonian.\cite{Schnyder2012PRB,ChiuTeoSchnyderRyu2016RMP} \rev{For our perpendicular Rashba and triplet spin textures the triplet contribution couples opposite helicities. We therefore compute the index using the full BdG matrix at finite pairing.}

\section{Charge-\texorpdfstring{$4e$}{4e} (quartet) condensates in quaternion language}\label{sec:4e}

Write the pair field as
\( q(\mathbf r)=\psi(\mathbf r)\,\mathbf 1+\bm d(\mathbf r)\!\cdot\!(\boldsymbol{e}_x,\boldsymbol{e}_y,\boldsymbol{e}_z) \),
with quaternion units obeying \(\boldsymbol{e}_i\boldsymbol{e}_j=-\delta_{ij}+\epsilon_{ijk}\boldsymbol{e}_k\) (Sec.~\ref{sec:prelim}).
The symmetric product of two pairs defines a \emph{quartet} (charge-\(4e\)) tensor,
\begin{equation}
\rev{\mathcal Q}(\mathbf r)\;\equiv\;\big\langle q(\mathbf r)\otimes q(\mathbf r)\big\rangle_{\rm sym}.
\label{eq:Q-def}
\end{equation}
Using quaternion multiplication,
\begin{align}
q^2
&=
\big(\psi^2-\bm d\!\cdot\!\bm d\big)\,\mathbf 1
+\big(2\psi\,\bm d\big)\!\cdot\!
(\boldsymbol{e}_x,\boldsymbol{e}_y,\boldsymbol{e}_z).
\label{eq:q-square}
\end{align}
Quaternion multiplication projects the symmetric tensor onto the following scalar and vector combinations; it does not retain all independent tensor channels:
\begin{subequations}\label{eq:Q-scalar-vector}
\begin{align}
Q_s(\mathbf r)&\;\propto\;\rev{\big\langle\mathrm{Sc}\!\big[q^2(\mathbf r)\big]\big\rangle},
\label{eq:Q-scalar}\\
\bm Q_v(\mathbf r)&\;\propto\;\rev{\big\langle 2\,\psi(\mathbf r)\,\bm d(\mathbf r)\big\rangle}.
\label{eq:Q-vector}
\end{align}
\end{subequations}
{There are two independent spin scalars, $\psi^2$ and $\bm d\cdot\bm d$. We use their difference $O=\mathrm{Sc}(q^2)=\det\Delta'$ as the composite coupled to $Q\equiv Q_s$. Its microscopic meaning follows by forming two pairs in the same two distinct single-particle modes $A,B$ (each with two spin states). For $P^\dagger=\sum_{ab}\Delta_{ab} c^\dagger_{Aa}c^\dagger_{Bb}$, fermion anticommutation gives
\[
 (P^\dagger)^2=-2\det\Delta\;
 c^\dagger_{A\uparrow}c^\dagger_{A\downarrow}
 c^\dagger_{B\uparrow}c^\dagger_{B\downarrow}.
\]
Here $\det\Delta=\det\Delta'$ since $\det(\ii\sigma_y)=1$. This contraction selects $O$ for the specified spin and mode structure. The sum $\det\varrho(q)=\psi^2+\bm d\cdot\bm d$ is a second allowed scalar. Before quaternion multiplication, the symmetric product of two
triplet amplitudes contains the spin-two tensor
\[
T_{ij}=d_i d_j-\frac{\delta_{ij}}{3}\bm d\cdot\bm d.
\]
Because $T_{ij}$ is symmetric and traceless,
$\sum_{ij}T_{ij}\boldsymbol e_i\boldsymbol e_j=0$.
The triplet contribution to $\mathrm{Sc}(q^2)$ is therefore
$-\bm d\cdot\bm d$.
The susceptibilities of these channels and the mixing of the two
scalars depend on microscopic interactions and orbital form factors.
We retain the determinant channel.}

{For a pair-density wave, write $q(\bm r)=q_+e^{\ii\bm K\cdot\bm r}+q_-e^{-\ii\bm K\cdot\bm r}$, where $q_\pm=(\psi_\pm,\bm d_\pm)$ denote the two amplitudes. For distinct harmonics $0,\pm2\bm K$, the uniform part of $O$ is $2(\psi_+\psi_- -\bm d_+\cdot\bm d_-)$. A singlet pair-density wave gives the familiar product $\Delta_{\bm K}\Delta_{-\bm K}$ up to normalization. Both scalar contractions agree in this limit, so this example alone leaves their relative weights for mixed pairing undetermined.\cite{BergFradkinKivelson2009NatPhys,Agterberg2020ARCM,FernandesFu2021PRL}}

Under \({\rm U}(1)\) gauge, \(q\mapsto e^{2\ii\phi}q\), hence
\begin{equation}
Q\mapsto e^{4\ii\phi}\,Q,\qquad D^{(4e)}_i=\partial_i-4\ii eA_i,
\label{eq:gauge-4e}
\end{equation}
i.e.\ \(Q\) carries charge \(4e\) and couples minimally via \(D^{(4e)}\). A pure-\(Q\) condensate therefore has halved flux quantum \(\Phi_0^{(4e)}=h/4e=\Phi_0/2\).
The selected scalar composite transforms as $\mathcal T:Q\mapsto Q^\ast$; a uniform phase can be chosen real in a time-reversal-invariant state. \cite{Ciaccia2024CommPhys}

Retaining the scalar \(Q\) and the pair quaternion \(q\), we consider
the following minimal local, gauge-invariant GL free energy:
\begin{subequations}\label{eq:GL-qQ}
\begin{align}
F[q,Q]&=\int d^d\mathbf r\ \big\{f_2+f_4+f_{\rm grad}+f_{\rm coup}\big\},\\
f_2&=\alpha\,|q|^2+r_Q\,|Q|^2,\label{eq:F2}\\
f_4&=\beta_1|q|^4+\beta_2\big|\mathrm{Sc}(q^2)\big|^2
+\beta_3|\bm N|^2+\lambda\,|Q|^4,\label{eq:F4}\\
f_{\rm grad}&=\kappa_1\,|(\boldsymbol\nabla-2ie\boldsymbol A)q|^2
+\eta\,|(\boldsymbol\nabla-4ie\boldsymbol A)Q|^2 \nonumber \\
&+\sum_{ij}\kappa_{ij}\,\mathrm{Sc}\!\left[(D_i q)(D_j q)^{\ddagger}\right], \label{eq:Fgrad}\\
f_{\rm coup}&=
g\,\mathrm{Re}\!\big[\,Q^\ast\,\mathrm{Sc}(q^2)\big]
+u\,|Q|^2|q|^2+\ldots .
\label{eq:Fcoup}
\end{align}
\end{subequations}

\rev{All coefficients in this functional are real.
The parameters $r_Q$, $\lambda$ and $\eta$ are the bare quartet
mass, quartic coefficient and gradient coefficient, respectively.
The coefficient $u$ couples the pair and quartet densities,
while $g$ couples $Q$ to the selected scalar composite.
The coefficients $\beta_1,\beta_2,\beta_3$ multiply the pair
quartic invariants specified in Sec.~\ref{sec:GL}.}

\rev{Equation~\eqref{eq:GL-qQ} uses the common pair mass and stiffness of Sec.~\ref{sec:GL}. The coupling $g$ multiplies the determinant channel defined above.}

Here \(D_i\equiv\partial_i-2\ii eA_i\), the quartet derivative is
\(D_i^{(4e)}\equiv\partial_i-4\ii eA_i\), and
\(\bm N=\ii\,\bm d\times\bm d^{\,\ast}\) is the \rev{triplet pair-spin} vector introduced in Sec.~\ref{sec:GL}.

To display the phase locking, write
\[
q(\mathbf r)=|q|\,\widehat q\,e^{\ii\varphi_q},
\qquad
Q(\mathbf r)=|Q|\,e^{\ii\varphi_Q},
\]
\rev{Here $|\widehat q|_H=1$, and $\widehat q$ specifies the
normalized internal singlet-triplet structure.}
Then
\[
\mathrm{Sc}\!\left[q^2\right]
=
|q|^2\,s(\widehat q)\,e^{2\ii\varphi_q},
\qquad
s(\widehat q)\equiv\mathrm{Sc}\!\left[\widehat q^{\,2}\right].
\]
\rev{For a nonzero $g\,s(\widehat q)$, choose the phase convention in which this product is real and}
\rev{negative. The trilinear term then reads}
\begin{equation}
f_{\rm coup}\supset
-|g|\,|Q|\,|q|^2\,|s(\widehat q)|\,
\cos\!\big(\varphi_Q-2\varphi_q\big).
\label{eq:phase-lock}
\end{equation}
\rev{The minimum in this convention satisfies}
\begin{equation}
\rev{\varphi_Q=2\varphi_q\pmod{2\pi}.}
\end{equation}
\rev{Locking requires $|q|\,|Q|>0$ and $g\,s(\widehat q)\ne0$. For $\psi=1$ and $\bm d=(0,0,1)$ the two contributions to $O$ cancel exactly, and this coupling leaves the relative phase free. In a general phase convention its preferred value is $\varphi_Q-2\varphi_q=\arg[-g\,s(\widehat q)]$ modulo $2\pi$.}\cite{SigristUeda1991}

{
The coefficient $\alpha(T)$ of $|q|^2$ in Eq.~\eqref{eq:F2}
is the bare charge-$2e$ pair mass.
For this fluctuation calculation we set $\bm A=0$ and
$\kappa_{ij}=0$.
Above the bare pair transition, $\alpha(T)>0$.
With $\kappa_1>0$, the four complex components
$a_\mu=(\psi,d_x,d_y,d_z)$, $\mu=0,1,2,3$, have the common
Gaussian propagator
\begin{align}
\big\langle a_\mu(\bm k)a_\nu^*(\bm k')\big\rangle
&=(2\pi)^d\delta^{(d)}(\bm k-\bm k')\delta_{\mu\nu}
  \frac{k_BT}{\alpha(T)+\kappa_1k^2}.
\label{eq:pair-gaussian-propagator}
\end{align}
}
For the unnormalized scalar projection $O=\mathrm{Sc}(q^2)=\psi^2-\bm d\cdot\bm d$ and the coupling $g\,\mathrm{Re}(Q^*O)$ in Eq.~\eqref{eq:Fcoup}, the leading quartet mass is
\begin{align}
r_{Q,\rm eff}
&=\bar r_Q-2g^2 k_B T\,\Pi(0),
\label{eq:mu-eff-def-main} \\
\bar r_Q&=r_Q+4u k_B T\,\mathcal I_0, \\
\mathcal I_0&=\int^{\Lambda}\!\frac{d^d k}{(2\pi)^d}
\frac{1}{\alpha+\kappa_1k^2}.
\end{align}
\rev{Here $\Lambda$ is the ultraviolet momentum cutoff.}
The $u$-term is a cutoff-dependent tadpole, which can be absorbed into $\bar r_Q$. The factor $2$ in the $g^2$ term follows from the four complex components and the explicit real-part convention. If all coefficients are instead defined in $F/(k_B T)$, the corresponding reduced-coefficient formula has no explicit $k_B T$ factor.

\rev{The bubble $\Pi(P)$ is evaluated at total momentum $P=0$
for a uniform quartet.
For $0<d<4$ it is ultraviolet convergent.
The infinite-cutoff continuum value below gives the leading
infrared behavior at finite cutoff $\Lambda$ when
$\alpha\ll\kappa_1\Lambda^2$:}
\begin{align}
\Pi(0)
&=
\int\!\frac{d^d\mathbf{k}}{(2\pi)^d}
\frac{1}{(\alpha+\kappa_1 k^2)^2}
\nonumber\\
&=
\frac{\Gamma\!\left(2-\tfrac{d}{2}\right)}
{(4\pi)^{d/2}}\,
\alpha^{\,\frac{d}{2}-2}\,
\kappa_1^{-\,\frac{d}{2}} .
\label{eq:Pi0-main}
\end{align}
In \(d=2,3\), this gives
\begin{subequations}\label{eq:Pi0-23-main}
\begin{align}
d=2: &\quad
\Pi(0)=
\frac{1}{4\pi}\,
\frac{1}{\alpha\,\kappa_1},
\label{eq:Pi0-2D-main}\\
d=3: &\quad
\Pi(0)=
\frac{1}{8\pi}\,
\frac{1}{\kappa_1^{3/2}\,\alpha^{1/2}} .
\label{eq:Pi0-3D-main}
\end{align}
\end{subequations}
\rev{The infrared singularity is proportional to $\alpha^{-1}$ in two dimensions and to $\alpha^{-1/2}$ in three. The prefactors have different units in the two dimensions.}

{A cutoff-independent test for a continuous preemptive transition uses the static uniform susceptibilities of the normal state. Let $\Gamma_4(T)$ be the inverse response to a source coupled linearly to the scalar field $Q$, and $\Gamma_{2,\min}(T)$ the smallest eigenvalue of the inverse pair susceptibility. At a continuous quartet transition on cooling,
\[
 \Gamma_4(T_4)=0,\qquad
 \Gamma_{2,\min}(T_4)>0 .
\]
When both normal-state masses soften monotonically on cooling, the quartet zero then lies above the putative pair zero of the continued normal state. The coefficients of Eq.~\eqref{eq:mu-eff-def-main} must be matched to the same microscopic or measured normal-state response upon changing the cutoff. The separate bare mass and tadpole have no invariant zero. With this source convention, the Gaussian curvature
$r_{Q,\rm eff}$ estimates $\Gamma_4$.
In the auxiliary-field construction of Appendix~\ref{app:analytic},
the susceptibilities of $Q$ and $O$ have the same pole but can
differ by a regular background. The zero of $r_{Q,\rm eff}$ defines a local instability temperature $T_4^{\rm loc}$, which can differ from the transition temperature $T_4$ once phase fluctuations are included.}

{A negative matched quartet mass signals a local instability of $Q=0$. If the effective coefficient $\lambda_{\rm eff}$ of $|Q|^4$ is positive, the quartic truncation gives $|Q|^2=-r_{Q,\rm eff}/(2\lambda_{\rm eff})$. In this background the smallest pair mass is $\alpha+u|Q|^2-|g||Q|$ and must remain positive for a locally stable pure-quartet solution. Phase coherence depends in addition on the quartet stiffness and phase fluctuations. Appendix~\ref{app:analytic} derives the Gaussian normalization and estimates $g$ for a specific weak-coupling model. In that model the formal onset lies on the scale where pair interactions correct the Gaussian result.}

\rev{The quartet gradient contribution
$F_{Q,\mathrm{grad}}=\int d^d r\,
\eta|(\boldsymbol\nabla-4\ii e\bm A)Q|^2$
gives the current
$\bm J_Q=-\delta F_{Q,\mathrm{grad}}/\delta\bm A$,}
\begin{equation}
\mathbf J_Q
=
8e\,\eta\,
\mathrm{Im}\!\big[
Q^\ast(\boldsymbol\nabla-4\ii e\boldsymbol A)Q
\big].
\label{eq:quartet-current}
\end{equation}
For uniform \(|Q|\), writing \(Q=|Q|e^{\ii\varphi_Q}\), one obtains
\begin{align}
\mathrm{Im}\!\big[
Q^\ast(\boldsymbol\nabla-4\ii e\boldsymbol A)Q
\big]
&=
|Q|^2\big(\boldsymbol\nabla\varphi_Q-4e\boldsymbol A\big),
\nonumber\\
\mathbf J_Q
&=
8e\eta |Q|^2
\big(\boldsymbol\nabla\varphi_Q-4e\boldsymbol A\big)
\nonumber\\
&=
-32e^2\eta |Q|^2
\left(
\boldsymbol A-\frac{1}{4e}\boldsymbol\nabla\varphi_Q
\right).
\label{eq:quartet-current-london}
\end{align}
On a contour sufficiently far from an isolated vortex that the equilibrium supercurrent vanishes, single-valuedness of the quartet phase gives
\begin{equation}
\oint \boldsymbol\nabla\varphi_Q\!\cdot d\boldsymbol\ell\;=\;2\pi n
\quad\Rightarrow\quad
\Phi\equiv\oint\mathbf A\!\cdot d\boldsymbol\ell\;=\;n\,\frac{h}{4e},
\label{eq:flux-quantization-4e}
\end{equation}
Here $h$ has been restored. This is \(h/4e\) flux quantization for a pure \(Q\) condensate; on a finite contour with nonzero current, the corresponding statement concerns fluxoid quantization. \rev{In the coexistence regime, for fixed internal pair structure and single-valued pair phase, the locked winding numbers obey $n_Q=2n_q$. Uncompensated winding of $\varphi_Q-2\varphi_q$ then requires a relative-phase domain wall. If the internal texture permits a half winding of $\varphi_q$, an accompanying full winding of $\varphi_Q$ can keep the relative phase locked. Confinement by this term requires finite stiffness and $g\,s(\widehat q)\ne0$ in the surrounding condensate. If the scalar projection vanishes, the trilinear coupling supplies no wall tension.}\cite{BergFradkinKivelson2009NatPhys,Agterberg2020ARCM}

{
For the weak link we assume that reversing both phase drops leaves
the junction energy invariant.
In a gauge with $\int_R^L\bm A\cdot d\bm\ell=0$, we retain
the pair-transfer harmonics $J_1,J_2$ and the direct
quartet-transfer amplitude $\widetilde J_1$,
\begin{align}
\mathcal E_J(\varphi,\varphi_4)
&=-J_1\cos\varphi-J_2\cos(2\varphi)\nonumber\\
&\quad-\widetilde J_1\cos\varphi_4+\cdots,
\label{eq:josephson-energy}\\
\varphi
&=\varphi_q^{(L)}-\varphi_q^{(R)},
\label{eq:josephson-phase}\\
\varphi_4
&=\varphi_Q^{(L)}-\varphi_Q^{(R)}.
\end{align}
For independent phase drops, direct quartet transfer contributes
$(4e/\hbar)\widetilde J_1\sin\varphi_4$ to the current.
When the locking offsets are equal in the two electrodes,
$\varphi_4=2\varphi$ modulo $2\pi$ and the second-harmonic
energy coefficient becomes
$J_{2,\mathrm{eff}}=J_2+\widetilde J_1$.
}

A dominant second harmonic $I(\varphi)=I_1\sin\varphi+I_2\sin(2\varphi)+\cdots$ \rev{can arise from coherent two-pair transfer}, but does not by itself demonstrate a bulk charge-$4e$ condensate. High-transparency junctions, spin-active barriers, $0$-$\pi$ cancellations, and SQUID interference can enhance the relative second harmonic in conventional superconducting devices. Doubled Josephson emission, half-spaced Shapiro positions, and half-period SQUID modulation can all follow from this same harmonic and are therefore not independent proofs of bulk quartet order.

Appendix~\ref{app:analytic} derives the ordinary weak-tunneling scaling: the first harmonic appears at second order in the single-electron tunneling amplitude and the coherent two-pair harmonic at fourth order. \rev{In a pure quartet condensate the junction phase drop obeys
$\dot\varphi_4=4eV/\hbar$, with $V$ oriented according to the
phase convention above.
Its fundamental oscillation frequency is $f_Q=4e|V|/h$.
A drive at frequency $f_{\rm d}$ permits locking at
$V_n=nhf_{\rm d}/(4e)$, $n\in\mathbb Z$.} In the planar SQUID of Ref.~\cite{Ciaccia2024CommPhys}, destructive interference suppresses the first harmonic and exposes coherent two-pair transport; this observation does not require a bulk quartet condensate.

\rev{Soldini, Fischer, and Neupert found quartet correlations in an attractive two-orbital chain. Their exact diagonalization of an eight-site ring gives four ground-state energy minima spaced by approximately $h/4e$ within one electronic flux period $h/e$. The relative values of $J_1$ and $J_2$ depend on the junction.}\cite{Soldini2024PRB}

\section{Model examples and numerical correlations}\label{sec:numerics}

\subsection{\rev{Topological transition in the Rashba lattice model}}

We study a Rashba metal on a square lattice with TRI pairing written directly
in quaternion variables. The normal-state electron Hamiltonian is
\begin{align}
h(\bm k)
&=
\big[-2t(\cos k_x+\cos k_y)-\mu\big]\Id
\nonumber\\
&\quad
+\alpha_R\big(\sin k_y\,\sigma_x-\sin k_x\,\sigma_y\big),
\label{eq:normal-rashba-model}
\end{align}
\rev{Here $t>0$ is the nearest-neighbour hopping,
$\mu$ is the chemical potential, $\alpha_R$ is the Rashba coupling,
and $\Id$ is the spin identity.
We set the lattice spacing to $a=1$, so $k_x,k_y$ are dimensionless.
With real amplitudes $\Delta_s$ and $\Delta_t$ for on-site
singlet and helical triplet pairing, respectively,
the gap quaternion is}
\begin{align}
q(\bm k)
=
\Delta_s
+
\Delta_t\big(\sin k_x\,\boldsymbol{e}_x+\sin k_y\,\boldsymbol{e}_y\big).
\label{eq:rashba-gap-quaternion}
\end{align}
In the absence of SOC, the compact quaternionic BdG form reduces to
\begin{align}
H_{\rm BdG}^{(0)}(\bm k)
=
\xi_{\bm k}\tau_z
+
\tau_+q_\Delta(\bm k)
+
\tau_-q_\Delta^{\ddagger}(\bm k).
\label{eq:BdG-compact-rashba-noSOC}
\end{align}
With Rashba SOC included, the normal-state spin-orbit term is retained in the
diagonal particle-hole blocks as
\begin{align}
H'_{\rm BdG}(\bm k)
&=
\begin{pmatrix}
h(\bm k) & \Delta'(\bm k)\\
\Delta'^\dagger(\bm k) & -\sigma_y h^{\mathsf T}(-\bm k)\sigma_y
\end{pmatrix},\nonumber\\
\Delta'(\bm k)&=\varrho\!\big(q_\Delta(\bm k)\big).
\label{eq:BdG-rashba-full}
\end{align}

For the representative set
\[
\begin{gathered}
t=1,\quad\mu=-2.0,\quad\alpha_R=0.6,\\
\Delta_s=0.25,\quad\Delta_t=0.35,
\end{gathered}
\]
the physical Hamiltonian is fully gapped and has class-DIII symmetries, but is topologically trivial. In this model $\bm g=\alpha_R(\sin k_y,-\sin k_x,0)$ is perpendicular to $\bm d=\Delta_t(\sin k_x,\sin k_y,0)$. Continuously reducing the triplet amplitude to zero leaves the gap open and connects the system to an ordinary singlet superconductor, as shown in Appendix~\ref{secS:edge}. Thus $\nu=0$ for these parameters; no protected helical edge crossing is inferred.

\begin{figure}[!tbp]
  \centering
  \includegraphics[width=0.92\linewidth]{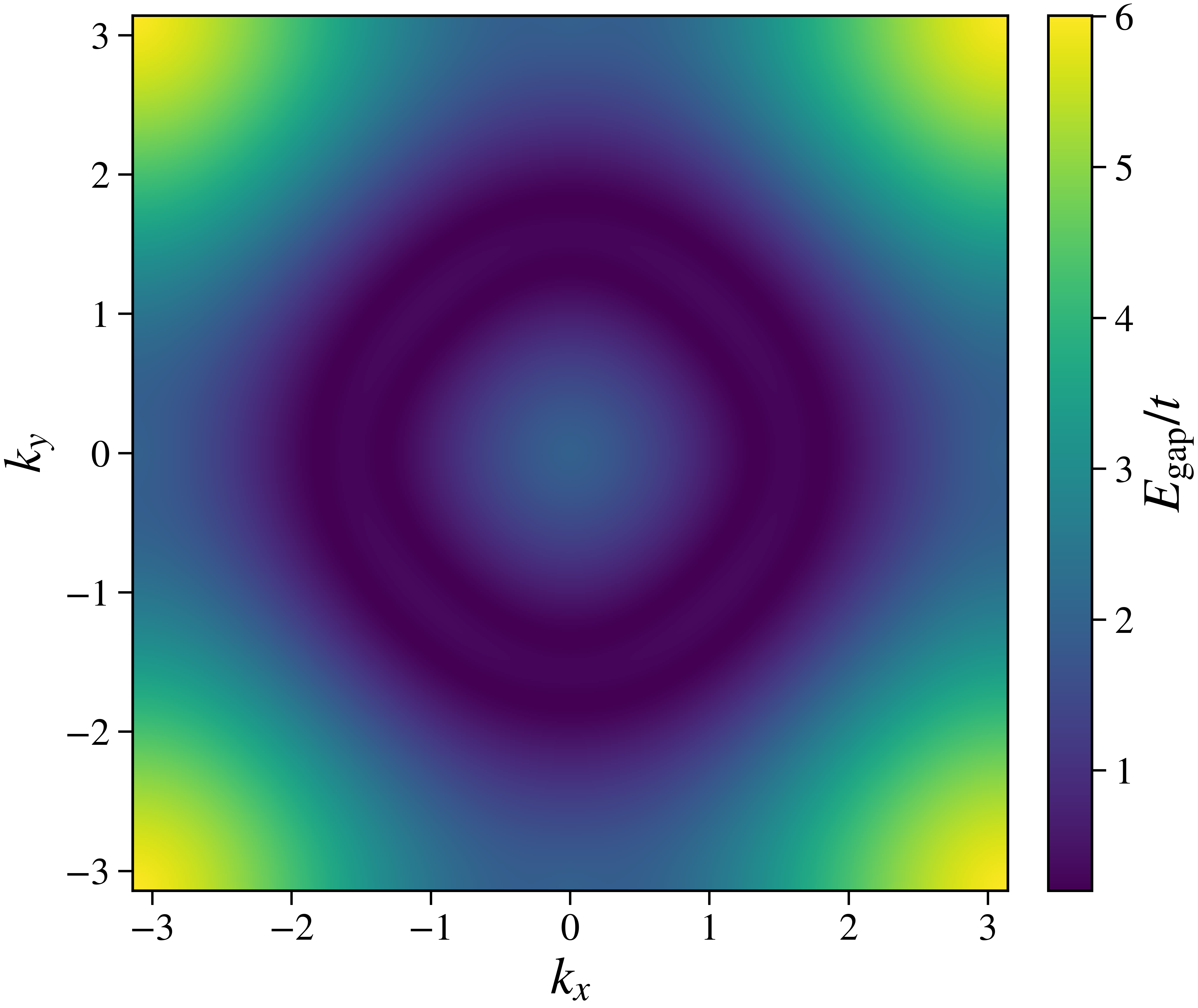}
  \caption{BdG excitation gap $E_{\rm gap}(\bm k)/t$ from Eq.~\eqref{eq:bdg-excitation-gap}, evaluated by diagonalizing the full spinful Hamiltonian on a $501\times501$ mesh for \rev{$\Delta_t/t=0.35$, $\Delta_s/t=0.25$, $\alpha_R/t=0.60$ and $\mu/t=-2$}. The spectrum is fully gapped; these parameters belong to the topologically trivial class-DIII phase established in Appendix~\ref{secS:edge}.}
  \label{fig:bdg-gapmap}
\end{figure}

Figure~\ref{fig:bdg-gapmap} displays the momentum-resolved excitation gap
\begin{equation}
E_{\rm gap}(\bm k)=\min_n\big|E_n(\bm k)\big|,
\label{eq:bdg-excitation-gap}
\end{equation}
where $E_n(\bm k)$ are the eigenvalues of the full Hamiltonian in Eq.~\eqref{eq:BdG-rashba-full}, including SOC and mixed-pairing branch splitting. The minimum on the $501\times501$ mesh is $E_{\rm gap}\simeq0.20306\,t$. The nonzero gap is consistent with the deformation argument; the topological classification follows from that argument rather than from the gap map alone.

\begin{figure}[!tbp]
 \centering
 \includegraphics[width=\linewidth]{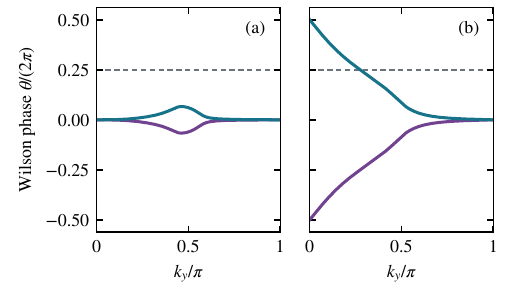}
 \caption{\rev{Wilson phases of the occupied BdG states for (a) $\Delta_t/t=0.35$ and (b) $\Delta_t/t=0.80$. Each loop contains 320 points, with 161 transverse momenta. The dashed reference at $\theta/(2\pi)=1/4$ has zero and one crossing, respectively. The endpoints $-1/2$ and $+1/2$ represent the same Wilson phase.}}
 \label{fig:rashba-wilson}
\end{figure}

{
Increasing only $\Delta_t/t$ to $0.80$ gives a second gapped phase. The excitation gap, $0.15000\,t$, occurs at $(\pm\pi/2,0)$ and $(0,\pm\pi/2)$. The normal dispersion, spin-orbit coupling, singlet amplitude and triplet spin texture are the same in both calculations. Appendix~\ref{secS:edge} shows that, for the positive triplet
amplitudes scanned here, the bulk is nodal for
$\sqrt{241/600}\leq\Delta_t/t\leq0.65$ and gapped on either side. This scan concerns a prescribed time-reversal-invariant pairing field.

To determine the index directly, let $V_j(k_y)$ be the $4\times2$ matrix of negative-energy eigenvectors of Eq.~\eqref{eq:BdG-rashba-full} at the $j$th point of a periodic $k_x$ loop. Write the singular-value decomposition of each overlap as
$V_j^\dagger V_{j+1}=L_j\Sigma_jR_j^\dagger$,
where $L_j,R_j$ are unitary and $\Sigma_j$ contains the singular values.
The overlaps are nonsingular on the meshes used here, so their
unitary polar factors are $L_jR_j^\dagger$.
The ordered product of these factors, including the closing link,
has eigenvalues $e^{\ii\theta_n(k_y)}$, $n=1,2$. A change of occupied-state basis conjugates the closed product and leaves its spectrum invariant.\cite{SoluyanovVanderbilt2011,YuQiBernevigFangDaiVanderbilt2011}

Figure~\ref{fig:rashba-wilson} shows the two phases for $0\leq k_y\leq\pi$. Counting crossings of $\theta/(2\pi)=1/4$ modulo one gives zero for $\Delta_t/t=0.35$ and one for $0.80$. Their parities are the strong indices $\nu=0$ and $\nu=1$. The counts agree on all three meshes, with $80\times41$, $160\times81$ and $320\times161$ points on the full $k_x$ loop and the half interval in $k_y$. In the nontrivial phase, a strip open along $y$ also exhibits a pair of counterpropagating modes on each boundary (Appendix~\ref{secS:edge}).}

\subsection{Illustrative charge-\texorpdfstring{$4e$}{4e} vortex}

\rev{In a locally stable pure-quartet regime the effective free energy contains $r_{Q,\rm eff}|Q|^2+\lambda_{\rm eff}|Q|^4+\eta|D^{(4e)}Q|^2$ with $\lambda_{\rm eff},\eta>0$. For $r_{Q,\rm eff}<0$ the homogeneous amplitude is $|Q|_\infty=\sqrt{-r_{Q,\rm eff}/(2\lambda_{\rm eff})}$ and the characteristic core scale is $\xi_Q\sim\sqrt{\eta/|r_{Q,\rm eff}|}$. The coefficients here are those of the effective theory after integrating out pair fluctuations.}

\begin{figure}[!tbp]
  \centering
  \begin{subfigure}[b]{0.49\linewidth}
    \centering
    \includegraphics[width=\linewidth]{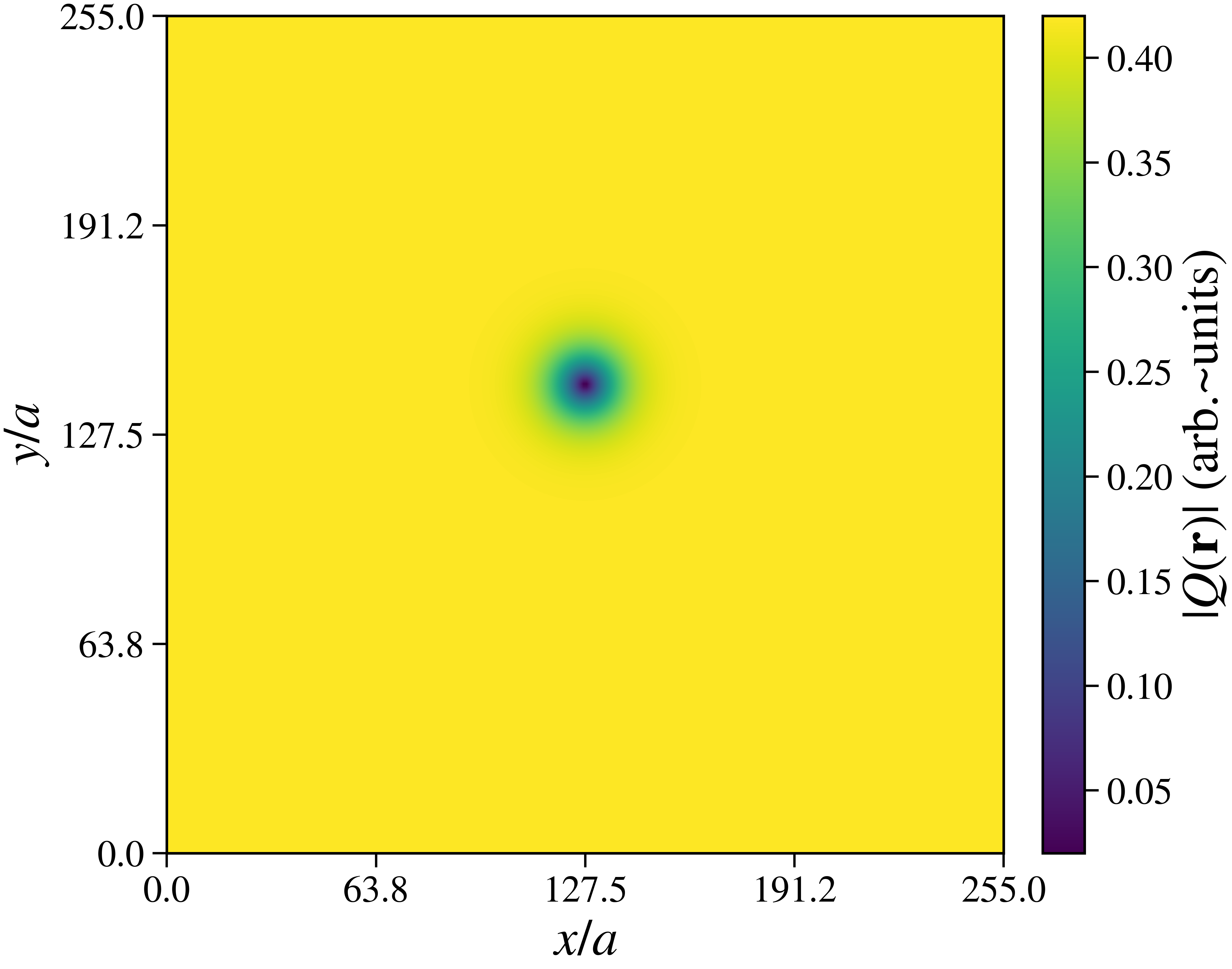}
    \caption{$|Q(\mathbf r)|$}
    \label{fig:Q-mag}
  \end{subfigure}\hfill
  \begin{subfigure}[b]{0.49\linewidth}
    \centering
    \includegraphics[width=\linewidth]{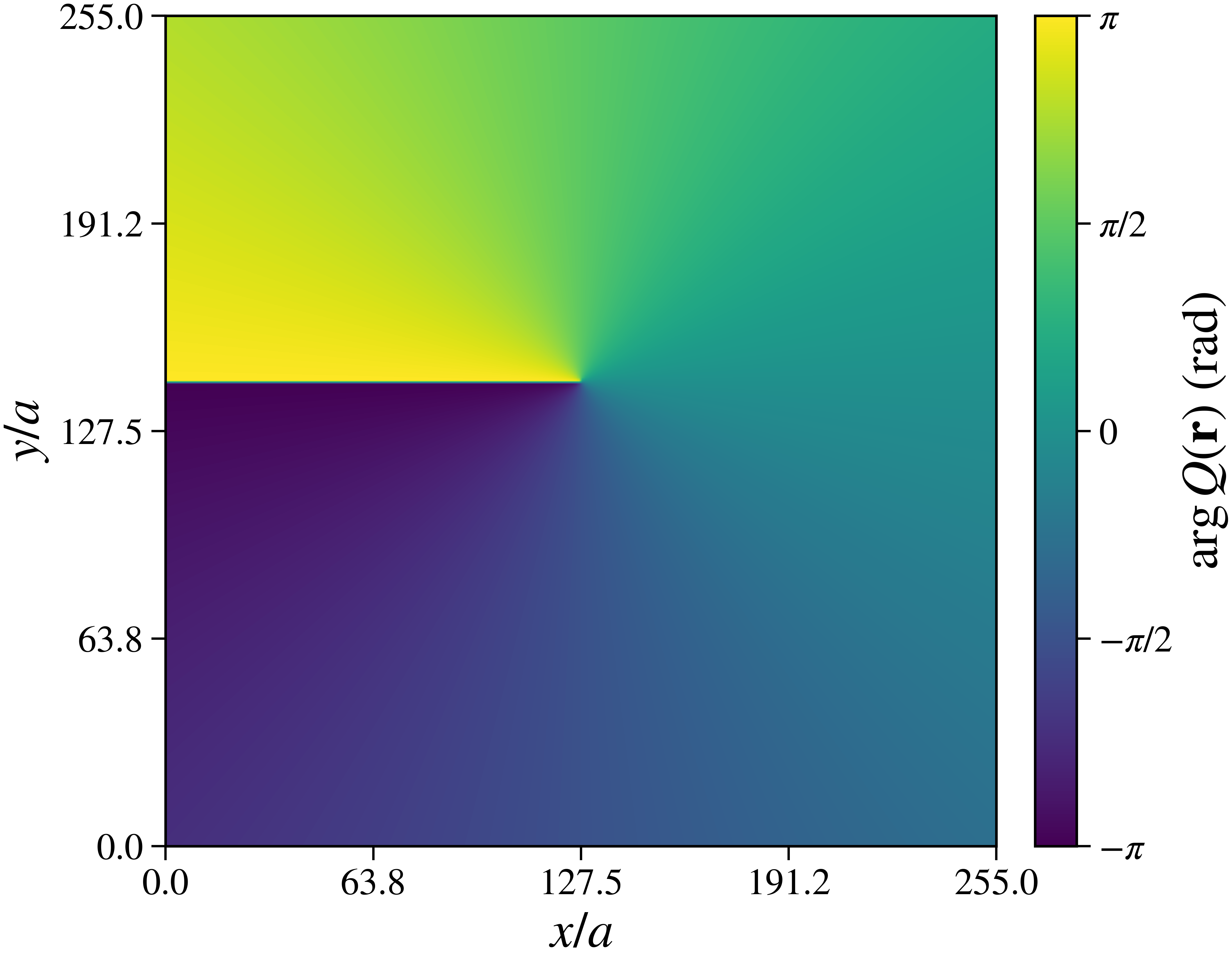}
    \caption{$\arg Q(\mathbf r)$}
    \label{fig:Q-phase}
  \end{subfigure}
  \caption{Prescribed quartet-vortex profile of Eq.~\eqref{eq:prescribed-vortex}: (a) amplitude and (b) phase with one $2\pi$ winding. Spatial coordinates are in lattice units ($a=1$); amplitude is in arbitrary units and phase in radians. The analytic $h/4e$ flux quantum follows from quartet gauge coupling and the asymptotic zero-current condition, rather than from a flux calculation in this graphic.}
  \label{fig:Q-vortex}
\end{figure}

Figure~\ref{fig:Q-vortex} illustrates a singly wound quartet field through the prescribed profile
\begin{equation}
Q(\bm r)=Q_0\tanh\!\left(\frac{r_c}{\sqrt{2}\,\xi}\right)e^{\ii\theta_c},
\label{eq:prescribed-vortex}
\end{equation}
where $r_c$ and $\theta_c$ are polar coordinates about the chosen core. The plotting parameters are $Q_0=0.42$, $\xi=8$, and a $256\times256$ grid with spacing $a=1$. They specify an illustrative amplitude and phase texture, not fitted GL coefficients or a self-consistent numerical minimization. The amplitude vanishes linearly near the core and tends to a uniform value far away. The $2\pi$ phase winding yields the analytic flux quantum $h/4e$ under the zero-current condition of Eq.~\eqref{eq:flux-quantization-4e}. The graphic does not include a computed vector potential or an independent numerical flux measurement.

\subsection{Charge-\texorpdfstring{$2e$}{2e} and charge-\texorpdfstring{$4e$}{4e} correlations in a 1D microscopic lattice model}
\label{sec:quartet-1d}

We study a two-orbital chain with interorbital attraction,
\begin{subequations}
\begin{align}
H &= \sum_{k,\sigma}
\boldsymbol c^\dagger_{k\sigma}
\big[\varepsilon(k)\,\rho_0-t_\perp\rho_x\big]
\boldsymbol c_{k\sigma}
-U\sum_i n_{i1}n_{i2}, \\
\boldsymbol c_{k\sigma}
&=(c_{k1\sigma},c_{k2\sigma})^\top, \\
\varepsilon(k)&=-2t\cos k-\mu .
\end{align}
\end{subequations}
\rev{For the open chain, $k=n\pi/(L+1)$ with
$n=1,\ldots,L$ and rung spacing set to one.
The operators are the standing-wave modes
\[
c_{ka\sigma}
=\sqrt{\frac{2}{L+1}}
\sum_{j=0}^{L-1}\sin[k(j+1)]\,c_{ja\sigma},
\]
where $j$ numbers the rungs from zero.
The nearest-neighbour hopping amplitude along each orbital is $t>0$.}
Here \(\rho_0\) is the identity and \(\rho_x\) is a Pauli matrix
in orbital space; \(t_\perp\) is the interorbital hybridization,
\(U>0\) is the interorbital attraction, \(\mu\) is the chemical
potential, and
\begin{equation}
n_{ia}=\sum_\sigma c^\dagger_{ia\sigma}c_{ia\sigma}
\end{equation}
is the density on orbital \(a=1,2\). The attractive density interaction favors interorbital occupancy and permits several pairing channels; here we measure the interorbital singlet channel. We define the local charge-\(2e\) pair operator
\begin{equation}
\Delta_i
=
c_{i1\uparrow}c_{i2\downarrow}
-
c_{i1\downarrow}c_{i2\uparrow},
\label{eq:Delta_i_1d}
\end{equation}
and the fully local charge-\(4e\) quartet operator
\begin{equation}
Q_i
=
c_{i1\uparrow}c_{i1\downarrow}
c_{i2\uparrow}c_{i2\downarrow}.
\label{eq:Q_i_1d}
\end{equation}
The corresponding equal-time correlators are
\begin{equation}
C_{2e}(r)=\langle \Delta_i^\dagger\Delta_{i+r}\rangle,
\qquad
C_{4e}(r)=\langle Q_i^\dagger Q_{i+r}\rangle .
\label{eq:C2e_C4e_def}
\end{equation}

\rev{In their grand-canonical two-orbital model, Soldini, Fischer and Neupert identified a one-dimensional quartet phase through algebraic four-body correlations and shorter-ranged one- and two-body correlations.\cite{Soldini2024PRB} Their Hamiltonian has intraorbital and interorbital attractions and orbital-conserving hopping. Here the intraorbital attraction is zero, $t_\perp=0.6t$, and the filling is fixed. We ask how the local quartet and interorbital singlet correlations compare over resolved distances in this chain. Appendix~\ref{secS:quartet_chain} applies the same slope analysis to two of their published iDMRG parameter sets.}

Ground states and correlation functions are computed with DMRG in its
matrix-product-state (MPS) formulation using \texttt{TeNPy}.\cite{White1992,White1993,Schollwock2011,HauschildPollmann2018TenPy}
We use open boundary conditions and system sizes up to \(L=80\) rungs, with four
fermionic modes per rung. Simulations are performed in the canonical ensemble
with \(U(1)\) particle-number conservation, fixing the total particle number to
\(N=2L\), corresponding to half filling. 
At fixed \(N\), the chemical-potential contribution \(-\mu N\) is a
constant and does not affect the ground-state wave function or these
correlation functions.
\rev{The archived correlators use a reference rung $i=L/2$ (zero-based indexing) and extend to $r=L/2-1$. Additional finite-chain calculations in Appendix~\ref{secS:quartet_chain} average over all rung pairs at each separation.}

The maximum MPS bond dimension $\chi_{\max}$ controls the number of retained Schmidt states. We compare $\chi_{\max}=800,1000,1200$, with $\chi_{\rm ref}=1200$ as the reference, and use the operational magnitude cutoff
\begin{equation}
|C_i(r)|\ge C_{\rm floor}=10^{-16},\qquad i\in\{2e,4e\},
\label{eq:dmrg-floor-main}
\end{equation}
for the pointwise relative-deviation analysis in Appendix~\ref{secS:quartet_chain}. This cutoff is not a universal floating-point accuracy bound and does not establish convergence by itself. At $L=80$, both channels exceed it through $r=15$ for all three bond dimensions. \rev{Between $\chi=800$ and $1200$, the pair and quartet magnitudes differ by $5.24\%$ and $10.00\%$ at $r=10$, and by $15.05\%$ and $23.14\%$ at $r=15$, respectively, relative to the $\chi=1200$ values.}

\rev{The fixed particle number gives $\langle\Delta_i\rangle=\langle Q_i\rangle=0$. We use the bare correlator magnitudes with the normalizations of Eqs.~\eqref{eq:Delta_i_1d} and \eqref{eq:Q_i_1d}.}

\rev{For channel $i\in\{2e,4e\}$, we fit $\ln|C_i(r)|$
against $\ln r$ by unweighted least squares over
$w=2h+1$ consecutive separations $r_{k-h},\ldots,r_{k+h}$.
Each window has its own intercept $A_i(r_k)$ and fitted
decay rate $\eta_i(r_k)$,}
{
\begin{equation}
\ln|C_i(r)|\approx A_i(r_k)-\eta_i(r_k)\ln r.
\label{eq:eta_fit_main}
\end{equation}
}
\rev{The fitted rate estimates the local logarithmic derivative
of a smooth decay curve at the window centre,}
{
\begin{equation}
\eta_i(r_k)\approx
-\left.\frac{d\ln|C_i(r)|}{d\ln r}\right|_{r=r_k}.
\label{eq:eta_def_main}
\end{equation}
}
For the representative plot, sliding windows contain seven consecutive points with $r\ge2$. A slope is retained only if every point in that channel's window satisfies
\begin{equation}
|C_i(r)|\ge C_\eta,\qquad C_\eta=10^{-14}.
\label{eq:C_floor_main}
\end{equation}
\rev{At $L=80$ the common window centres are $r=5,\ldots,10$ for both $\chi=800$ and $1200$. The seven-point fits therefore sample separations from 2 to 13. For an exponential $|C(r)|\propto e^{-r/\xi}$ the differential slope is $r/\xi$. An unweighted fit of $\ln|C_i|$ against $r$ over these same twelve separations at $\chi=1200$ gives effective exponential lengths of 0.429 and 0.463 rungs for the pair and quartet, respectively. These fits describe the measured interval and leave the asymptotic decay open.}

To reduce the influence of the shortest distances, the supplied plotting analysis uses
\begin{equation}
S_i^{\rm tail}(r_0)=\frac1L\sum_{r=r_0}^{r_{\max}}(L-r)|C_i(r)|,
\label{eq:Stail_def_main}
\end{equation}
with $r_{\max}=39$ for the $L=80$ data, and the ratio
\begin{equation}
R_{\rm tail}(L;r_0)=\frac{S_{4e}^{\rm tail}(r_0)}{S_{2e}^{\rm tail}(r_0)}.
\label{eq:Rtail_def_main}
\end{equation}
These plotted sums include all recorded separations; the curve is omitted when the denominator falls below $10^{-18}$. This denominator threshold differs from both $C_\eta$ and the pointwise cutoff in Eq.~\eqref{eq:dmrg-floor-main}. Since the data use a single central reference rung, $(L-r)$ is a chosen distance weight, not an explicit average over all rung pairs at separation $r$.

\begin{figure}[!tbp]
  \centering
  \includegraphics[width=\linewidth]{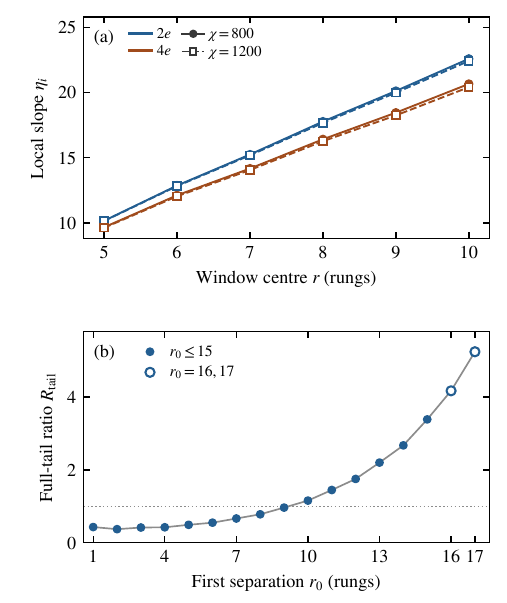}
  \caption{\rev{Correlations at $L=80$, $N=2L$,
$t_\perp/t=0.6$, $U/t=4$ and $\mu/t=-1.2$
for a central reference rung.
(a) Seven-point log-log slopes at the six common window centres
$r=5,\ldots,10$ for $\chi=800$ and $1200$.
Every fitted magnitude satisfies $|C_i|\ge10^{-14}$.
(b) Full-tail ratio at $\chi=800$, summed through $r=39$
and shown while $S_{2e}^{\rm tail}\ge10^{-18}$.
For filled circles ($r_0\le15$), both starting magnitudes exceed $10^{-16}$.
For open circles ($r_0=16,17$), both start below it.
The dotted line marks unity for the stated operator normalizations.}}
  \label{fig:eta-rtail-1d}
\end{figure}

\rev{The quartet has the smaller local slope at all six common window centres in Fig.~\ref{fig:eta-rtail-1d}(a). For $\chi=1200$, $(\eta_{2e},\eta_{4e})$ rises from $(10.15,9.62)$ at $r=5$ to $(22.42,20.41)$ at $r=10$. In panel (b), the full-tail ratios starting at $r_0=16$ and 17 already begin below the common pointwise cutoff. Even the earlier-starting full tails include later, subthreshold terms. If both sums are restricted to $10\leq r\leq15$, where both channels pass at all three bond dimensions, the ratios are $1.157$, $1.211$ and $1.220$ for $\chi=800,1000,1200$. Rescaling either operator changes the absolute tail ratio.
On a fixed fitting window, the rescaling shifts $\ln|C_i|$
by a constant and leaves the slope unchanged.
The absolute cutoff $C_\eta$ can change which windows are retained.}

\subsection{Josephson CPR and doubled-frequency response}
\begin{figure}[!tbp]
  \centering
  \begin{subfigure}[b]{0.49\linewidth}
    \centering
    \includegraphics[width=\linewidth]{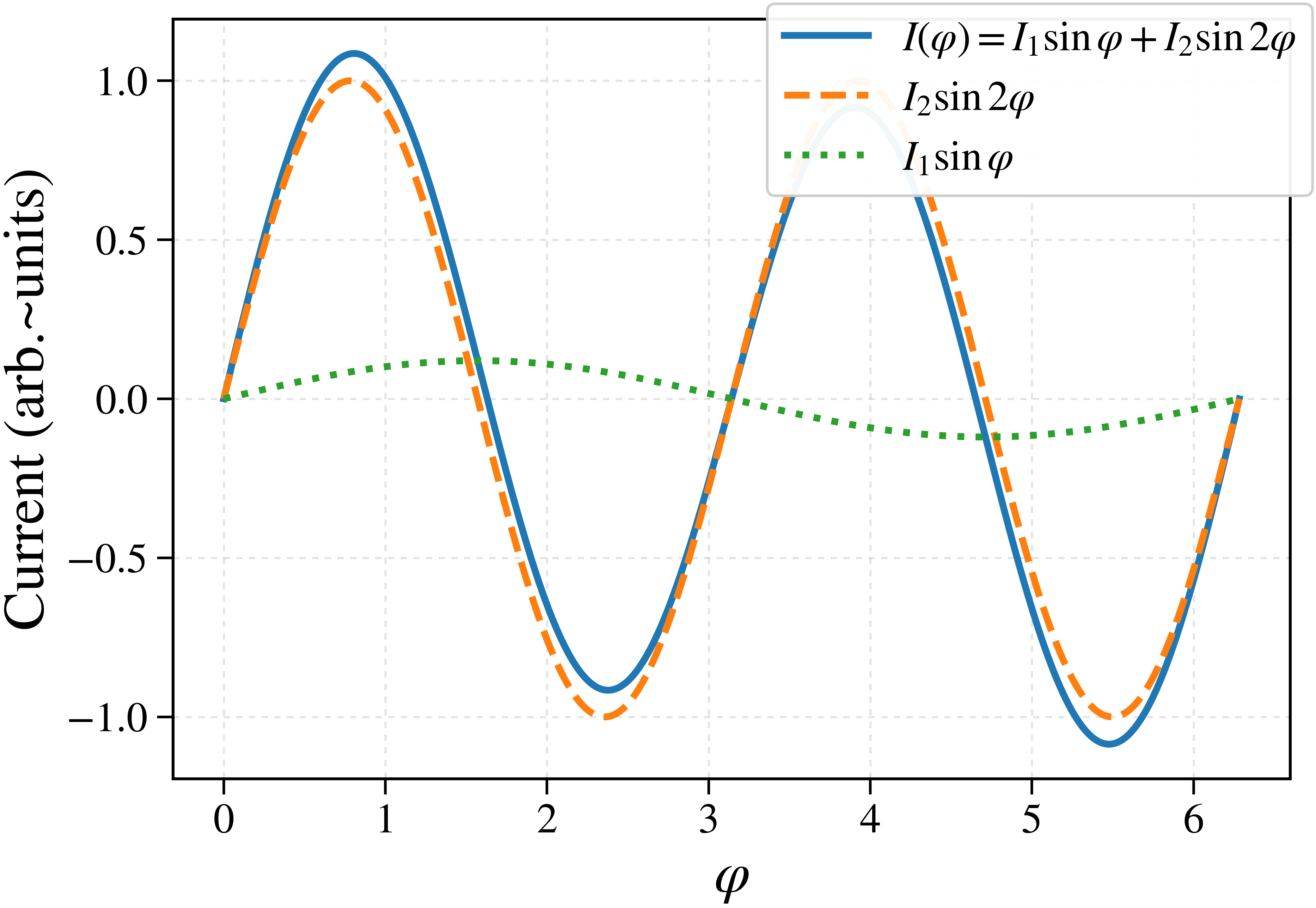}
    \caption{}
    \label{fig:CPR}
  \end{subfigure}\hfill
  \begin{subfigure}[b]{0.49\linewidth}
    \centering
    \includegraphics[width=\linewidth]{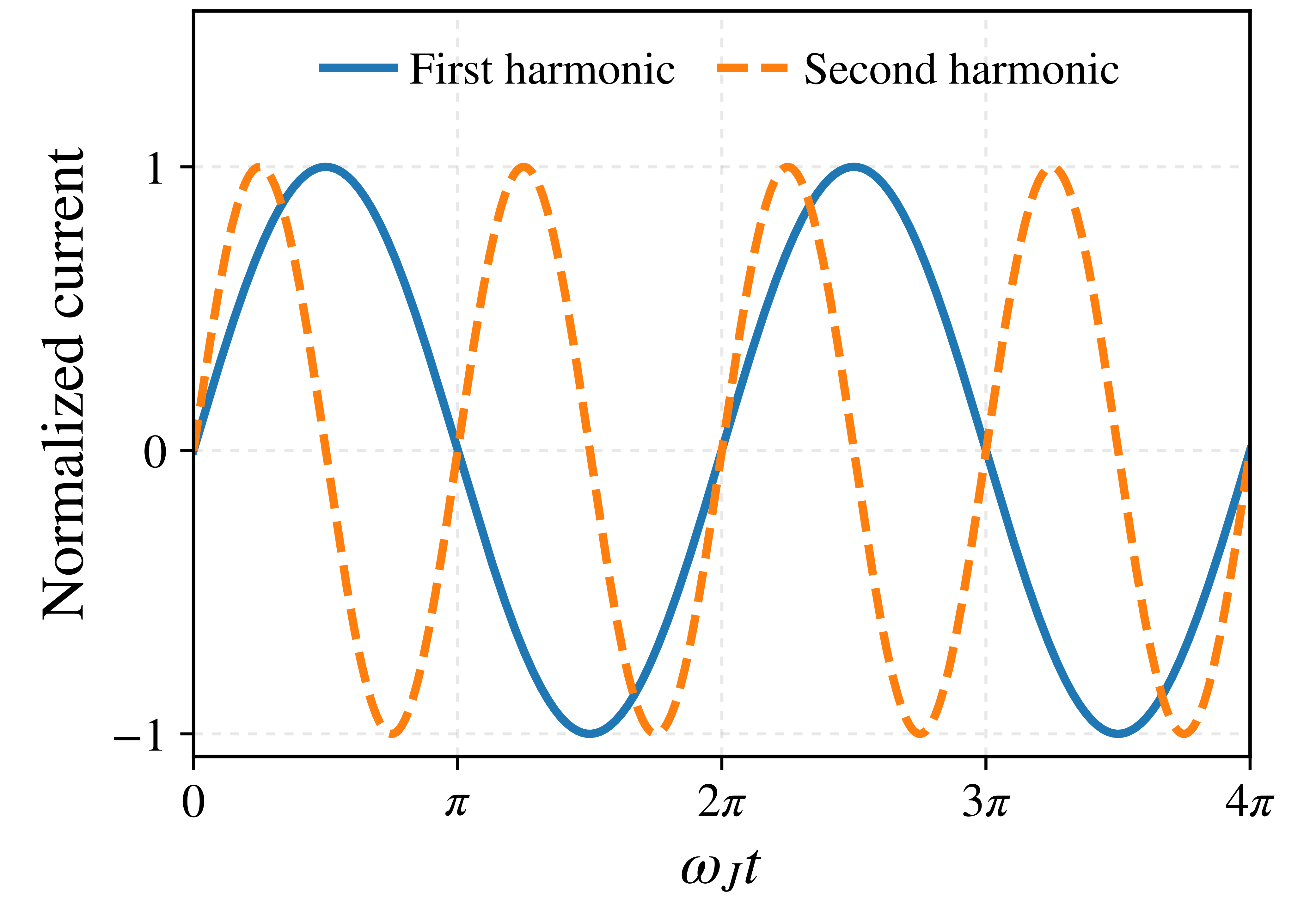}
    \caption{}
    \label{fig:AC}
  \end{subfigure}
  \caption{\rev{(a) Current-phase relation from the boundary energy in Eq.~\eqref{eq:josephson-energy} and its derivative in Eq.~\eqref{eq:CPR}. The total current and its two terms are shown for $I_1=0.12$ and $I_2=1$ in the same arbitrary current units. (b) Normalized currents for each harmonic separately under a DC voltage $V$. On the same time axis the second harmonic completes two cycles for every cycle of the first, with $\omega_J=2eV/\hbar$.}}
  \label{fig:CPR-AC}
\end{figure}

\rev{For independent phase drops, the pair-transfer current is
$I(\varphi)=(2e/\hbar)\partial_\varphi\mathcal E_J$
at fixed $\varphi_4$.
With Eqs.~\eqref{eq:josephson-energy}
and \eqref{eq:josephson-phase}, this gives}
\begin{align}
I(\varphi)&=\frac{2e}{\hbar}\big[J_1\sin\varphi+2J_2\sin(2\varphi)+\cdots\big]\nonumber\\
&\equiv I_1\sin\varphi+I_2\sin(2\varphi)+\cdots,
\label{eq:CPR}
\end{align}
\rev{Here $I_1=2eJ_1/\hbar$ and $I_2=4eJ_2/\hbar$.
For compatible locked phases, the total current has the same form
with $I_2=4eJ_{2,\mathrm{eff}}/\hbar$.} Figure~\ref{fig:CPR-AC}(a) uses the chosen amplitudes $I_1=0.12$ and $I_2=1$ in arbitrary units. \rev{We choose these amplitudes to illustrate a junction with a strong second harmonic. Their ratio depends on the tunneling matrix elements and the electrodes.}

\rev{A DC voltage gives $\dot\varphi=\omega_J=2eV/\hbar$.
The first and second current harmonics have frequencies
$f_J=|\omega_J|/(2\pi)=2e|V|/h$ and $2f_J$, respectively.} Figure~\ref{fig:CPR-AC}(b) compares normalized $\sin(\omega_Jt)$ and $\sin(2\omega_Jt)$ traces. When the second harmonic dominates, the largest oscillatory signal occurs at $2f_J$ even though the usual pair-phase Josephson relation is unchanged.

\rev{In the ordinary tunnel limit, scale the normal-state channel
transparencies as $\tau_a=c_a\tau$, with fixed dimensionless
$c_a$ and $\tau\ll1$.
At fixed temperature and with a nonzero leading first harmonic,
Appendix~\ref{app:analytic} gives
$J_1=O(\tau)$ and $J_2=O(\tau^2)$, hence}
\begin{equation}
\frac{J_2}{J_1}=\kappa_J(T,\text{junction parameters})\,\tau+O(\tau^2).
\label{eq:J2J1-scaling}
\end{equation}
The sign and magnitude of $\kappa_J$ depend on the channel structure, temperature, and barrier; the scaling alone supplies no universal numerical ratio. Higher harmonics at large transparency require a calculation beyond this expansion.\cite{Golubov2004RMP}

For identical electrodes with pair and quartet amplitudes, a useful boundary parametrization is
\begin{equation}
J_1=A\tau|q|^2,\qquad
J_{2,\rm eff}=B\tau^2|q|^4+C|Q|^2,
\label{eq:J1J2-phenom}
\end{equation}
where $A,B,C$ are junction-dependent coefficients with the required units. \rev{For the compatible locked phases of Sec.~\ref{sec:4e},
$C|Q|^2=\widetilde J_1$.
The coefficient $C$ depends on transmission through the contact.} Consequently,
\begin{equation}
\frac{J_{2,\rm eff}}{J_1}
=\frac BA\tau|q|^2+\frac{C}{A\tau}\frac{|Q|^2}{|q|^2}.
\label{eq:J2J1-effective}
\end{equation}
\rev{With $q\ne0$, a sufficient condition for direct quartet tunneling to dominate both harmonic contributions from pair tunneling is}
\begin{equation}
\rev{|C|\,|Q|^2\gg\max\!\left\{|A|\tau|q|^2,\ |B|\tau^2|q|^4\right\},}
\label{eq:quartet-window}
\end{equation}
subject to a stable quartet state and a coherent weak link. No transparency interval follows without specifying the boundary coefficients. In a pure-$Q$ state the pair phase is undefined; the direct quartet phase difference and its $4eV/\hbar$ Josephson relation should be used instead of this ratio.

These examples illustrate the harmonic response for specified junction coefficients; the distinction between coherent two-pair transport and bulk quartet order is discussed in Sec.~\ref{sec:4e}.

\subsection{SQUID periodicity and Shapiro steps}
\begin{figure}[!tbp]
  \centering
  \includegraphics[width=0.92\linewidth]{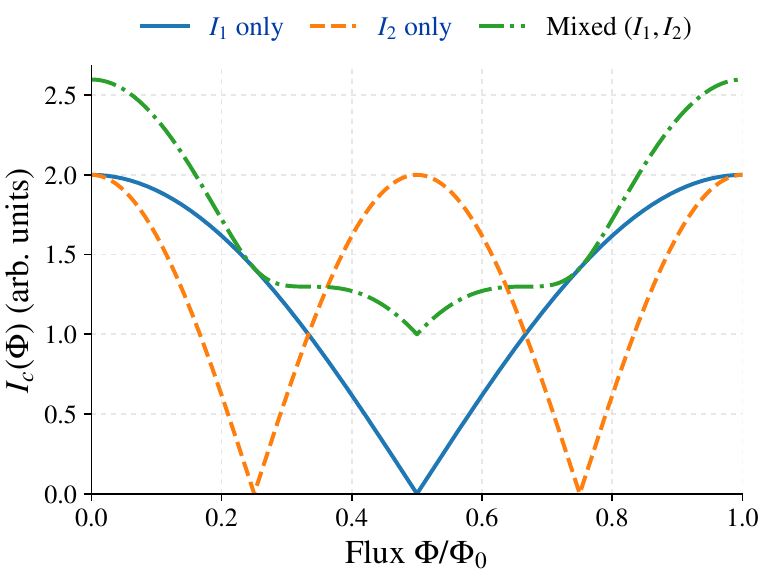}
  \caption{\rev{SQUID critical current for the first harmonic alone, the second harmonic alone, and their superposition. The three curves use $(I_1,I_2)=(1,0)$, $(0,1)$, and $(1,0.5)$, respectively.}
The flux is plotted in normalized units $\Phi/\Phi_0$ (dimensionless), with $\Phi_0\equiv h/2e$. For $I_1=0$ the modulation period halves to $\Phi_0/2=h/4e$, as follows directly from Eq.~\eqref{eq:squid-current}. The overall scale of $I_c$ is shown in arbitrary units set by the CPR amplitudes $I_1$ and $I_2$.}
  \label{fig:SQUID-Ic}
\end{figure}

For a symmetric two-junction SQUID with negligible loop inductance, write the junction phases as $\varphi\pm\delta$, where $\delta=\pi\Phi/\Phi_0$ and $\Phi_0=h/2e$. The total current is
\begin{equation}
I_{\rm SQ}=2I_1\sin\varphi\cos\delta
+2I_2\sin(2\varphi)\cos(2\delta),
\label{eq:squid-current}
\end{equation}
and $I_c(\Phi)=\max_\varphi|I_{\rm SQ}|$. Figure~\ref{fig:SQUID-Ic} compares $(I_1,I_2)=(1,0),(0,1),(1,0.5)$. The pure second-harmonic limit has period $\Phi_0/2$. The half-period follows from the second harmonic, irrespective of its microscopic origin.\cite{ClarkeBraginskiSQUID,Ciaccia2024CommPhys}

\begin{figure}[!tbp]
  \centering
  \includegraphics[width=\linewidth]{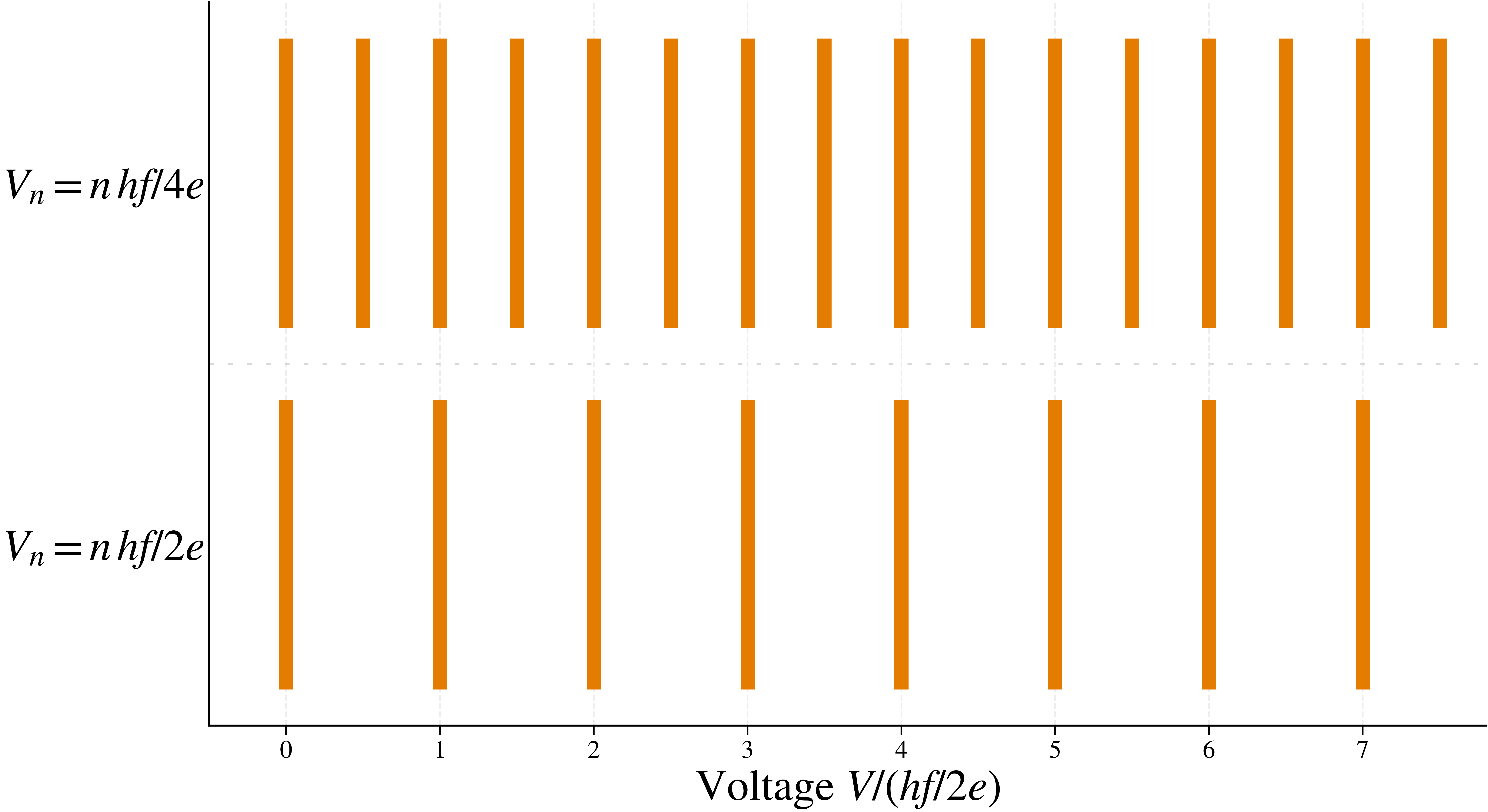}
  \caption{Schematic Shapiro locking positions at fixed drive frequency $f$. \rev{A junction with the first harmonic alone has candidate positions at $V_n=nhf/(2e)$ (bottom). For the second harmonic alone the spacing is $hf/(4e)$ (top), as also occurs for direct tunneling between phase-coherent quartet condensates (Sec.~\ref{sec:4e}). The horizontal axis gives voltage in units of $hf/(2e)$.}}
  \label{fig:Shapiro}
\end{figure}

Figure~\ref{fig:Shapiro} marks the candidate phase-locking voltages under a microwave drive of frequency $f$. A first-harmonic current locks at $V_n=nhf/(2e)$, whereas a pure second harmonic permits $V_n=nhf/(4e)$. The graphic is a schematic of these positions, not a dynamical microwave-response calculation. Step widths, missing steps, and stability depend on the driven junction dynamics. Their interpretation follows the discussion in Sec.~\ref{sec:4e}.\cite{Golubov2004RMP,Ciaccia2024CommPhys}

\section{Conclusions}\label{sec:conclusions}
\rev{In mixed singlet-triplet pairing the determinant of the physical gap is $\mathrm{Sc}(q^2)=\psi^2-\bm d\cdot\bm d$. Its magnitude fixes the product of the two pairing singular values. The full vector $\bm m$ fixes the splitting of the squared
singular values, $\lambda_+-\lambda_-=2|\bm m|$,
including the singlet-triplet interference absent from the
triplet pair-spin vector $\bm N$. Selecting the determinant as the quartet channel connects these gap properties to the phase-sensitive coupling. In coexistence, this term locks the relative phase when $g\,\mathrm{Sc}(\widehat q^{\,2})\ne0$.}

\rev{A negative renormalized quartet mass signals a local instability of the normal state. Within the quartic expansion, a locally stable pure-quartet solution also requires a positive effective quartet quartic and a positive smallest pair mass, $\alpha+u|Q|^2-|g||Q|$. Phase coherence depends separately on the stiffness and phase fluctuations. For four degenerate complex pair components, the Gaussian correction from the trilinear vertex is $-2g^2 k_BT\Pi(0)$. In the weak-coupling example of Appendix~\ref{app:analytic}, the resulting mass zero occurs on the scale of corrections from the pair interaction. Establishing a stable weak-coupling quartet interval would require treating those corrections and the quartet self-interaction.}

\rev{In the Rashba model, increasing $\Delta_t/t$ from $0.35$ to $0.80$ changes the DIII index from 0 to 1 through a finite nodal interval. The nontrivial phase has a bulk excitation gap of $0.15\,t$ and a helical edge pair. The Wilson-loop index characterizes the occupied states of the charge-$2e$ BdG model. Whether its pair fluctuations stabilize a charge-$4e$ condensate is a separate question governed by the quartet susceptibility and self-interaction. In the $L=80$ open chain the quartet slope lies below the singlet slope over window centres $r=5$-$10$ at both $\chi_{\max}=800$ and 1200. Both slopes increase across this interval, which leaves the asymptotic decay undetermined. The prescribed vortex illustrates a phase winding in a pure quartet field. For a weak link, coherent transfer of two Cooper pairs can itself give a half-period SQUID response. A half period alone leaves the nature of the bulk condensate open.}

\section*{Data availability}

\rev{The DMRG data and figure-generation scripts are available in the public GitHub repository
\url{https://github.com/Christian48596/Quaternionic-Superconductivity}.}


\appendix

\section{Fluctuation bubble and Josephson tunneling}
\label{app:analytic}

\subsection{Gaussian evaluation of \texorpdfstring{$\Pi(0)$}{Pi(0)}}

Near \(T_c\), in the classical static GL regime, the Cooper-pair field \(q\)
is treated as a bosonic order-parameter fluctuation. The microscopic fermions
have already been integrated out in obtaining the GL functional. Thus the
calculation below is not a Grassmann integral over electron fields, but a
Gaussian functional integral over the complex bosonic field \(q\).

Write the four complex amplitudes as $a_\mu=(\psi,d_x,d_y,d_z)$ and define $s_\mu=(1,-1,-1,-1)$. The scalar projection used throughout is $O=\mathrm{Sc}(q^2)=\sum_{\mu=0}^3s_\mu a_\mu^2$, with no additional normalization factor.
\rev{In the zero-field isotropic limit of
Eq.~\eqref{eq:pair-gaussian-propagator},
the quadratic pair free energy above the bare transition is}
\begin{equation}
F_q^{(2)}=\sum_{\mu=0}^3\int\!\frac{d^d k}{(2\pi)^d}
(\alpha+\kappa_1k^2)|a_\mu(\bm k)|^2,
\label{eqS:Fq2}
\end{equation}
where $\alpha,\kappa_1>0$.\cite{deGennes1966,TinkhamBook}
\rev{The covariance is given in Eq.~\eqref{eq:pair-gaussian-propagator}.
We separate its thermal factor by defining}
\begin{equation}
G_q(\bm k)=\frac{1}{\alpha+\kappa_1 k^2}.
\label{eqS:Gq}
\end{equation}
\rev{Thus the physical covariance contains $k_BT G_q$.}

The bubble \(\Pi(P)\) is the convolution of two thermal-factor-stripped propagators at total center-of-mass momentum \(P\). Since the vestigial
quartet considered in the main text is spatially uniform, the relevant
channel is \(P=0\). Physically, this describes a zero-momentum charge-\(4e\)
fluctuation resolving into two Cooper-pair fluctuations with opposite momenta
\(\mathbf k\) and \(-\mathbf k\). Therefore
\begin{equation}
\Pi(0)
=
\int \!\frac{d^d \mathbf{k}}{(2\pi)^d}\,
G_q(\mathbf k)G_q(-\mathbf k).
\label{eqS:Pi0-GG}
\end{equation}
\rev{Since $G_q(\bm k)=G_q(-\bm k)$, this is the integral
in Eq.~\eqref{eq:Pi0-main}. Its evaluation uses}
\begin{equation}
\int\!\frac{d^d\bm k}{(2\pi)^d}\frac{1}{(A+B k^2)^2}
=\frac{\Gamma(2-d/2)}{(4\pi)^{d/2}}
 A^{d/2-2}B^{-d/2},
\end{equation}
\rev{where $A,B>0$, $0<d<4$, and $\Gamma$ is the Euler gamma
function. Setting $A=\alpha$ and $B=\kappa_1$ gives
Eq.~\eqref{eq:Pi0-main}; its two- and three-dimensional
limits are Eqs.~\eqref{eq:Pi0-2D-main} and
\eqref{eq:Pi0-3D-main}.}

{
For uniform $Q$, we retain Gaussian pair fluctuations and absorb
the leading $u$ tadpole into the quartet mass.
The truncated functional used to calculate the $|Q|^2$ term is
\begin{equation}
F_{\rm tr}[q,Q]
=F_0[Q]+F_q^{(2)}[q]+F_{\rm coup}[q,Q].
\label{eqS:Fsplit}
\end{equation}
}
\begin{equation}
\begin{aligned}
F_0[Q]&=V\bar r_Q|Q|^2,\\
\bar r_Q&=r_Q+4u k_BT\!\int^\Lambda\frac{d^d k}{(2\pi)^d}G_q(\bm k).
\end{aligned}
\label{eqS:F0Q}
\end{equation}
\rev{In this subsection, $V=\int d^d r$ denotes the
spatial volume, or area in two dimensions.}
The leading $u|Q|^2|q|^2$ tadpole has been absorbed into $\bar r_Q$; it depends on the ultraviolet regularization. Retaining the $g$-vertex explicitly gives
\begin{equation}
F_{\rm coup}=\frac g2\int d^d r\,[Q^*O(\bm r)+QO^*(\bm r)].
\label{eqS:Fcoup}
\end{equation}
The normalized Gaussian average defines
\begin{equation}
F_{\rm eff}[Q]=F_0[Q]-k_BT\ln\left\langle
e^{-F_{\rm coup}/(k_BT)}\right\rangle_q.
\label{eqS:Feff-def}
\end{equation}
Since $\langle O\rangle_q=0$, the second cumulant is
\begin{equation}
F_{\rm eff}[Q]=F_0[Q]-\frac{1}{2k_BT}
\langle F_{\rm coup}^2\rangle_q+O(g^4).
\label{eqS:Feff-second}
\end{equation}
Odd cumulants vanish by charge neutrality in this Gaussian ensemble. Wick contraction gives, with $G_q(\bm r)$ the Fourier transform of $G_q(\bm k)$,
\begin{align}
\langle O(\bm r)O^*(\bm r')\rangle_q
&=2\sum_{\mu=0}^3|s_\mu|^2(k_BT)^2G_q(\bm r-\bm r')^2\nonumber\\
&=8(k_BT)^2G_q(\bm r-\bm r')^2.
\label{eqS:Scq2-corr}
\end{align}
Each complex component has two pairings, and there are four components. The two neutral cross terms in $F_{\rm coup}^2$ then yield
\begin{equation}
\langle F_{\rm coup}^2\rangle_q
=4g^2(k_BT)^2V|Q|^2\Pi(0).
\label{eqS:Fcoup2}
\end{equation}
Consequently,
\begin{equation}
F_{\rm eff}[Q]=V[\bar r_Q-2g^2k_BT\Pi(0)]|Q|^2+\cdots,
\label{eqS:Feff-mass}
\end{equation}
\begin{equation}
r_{Q,\rm eff}=\bar r_Q-2g^2k_BT\Pi(0).
\label{eqS:mu-eff}
\end{equation}
This coefficient is fixed by the stated normalization of $q$, $O$, and $g$. Equivalently, introducing $g_{\rm fl}=2g$ writes the correction as $-g_{\rm fl}^2k_BT\Pi(0)/2$; using that form with the original $g$ would miss the component factor. For coefficients normalized in $F/(k_BT)$, the explicit thermal factor disappears after all coefficients are rescaled consistently.

{At finite uniform $Q$, real and imaginary pair fluctuations in each complex component have quadratic coefficients $\alpha+u|Q|^2\pm|g||Q|$ in axes aligned with the phase of $Q$. The smaller coefficient is the stability test used in Sec.~\ref{sec:4e}.}

{\emph{Weak-coupling estimate.} Consider two spin-degenerate Fermi pockets exchanged by inversion. Take a constant singlet form factor and a triplet form factor equal to $+1$ and $-1$ on the two pockets, respectively. Tune the pairing interactions to equal singlet and triplet bare transition temperatures $T_c$ as in the common-mass model. Let $N_F$ denote the normal density of states per spin summed over both pockets, and take it constant over the pairing scale. With $\psi$ and $\bm d$ in energy units, expansion of the BCS fermion determinant through fourth order yields\cite{deGennes1966,TinkhamBook}
\[
 \begin{aligned}
 f_4^{\rm BCS}
 &=\frac{b}{2}\left\langle\mathrm{Tr}
       [(\Delta'\Delta'^\dagger)^2]\right\rangle_F
   =2b|q|^4-b|O|^2 ,\\
 b&=\frac{N_F k_BT_c}{2}\sum_{n=-\infty}^{\infty}
       \int_{-\infty}^{\infty}\frac{d\xi}{(\omega_n^2+\xi^2)^2}
   =\frac{7\zeta(3)N_F}{16\pi^2(k_BT_c)^2}.
 \end{aligned}
\]
Here $\omega_n=(2n+1)\pi k_BT_c$, $\xi$ is the normal-state energy relative to the chemical potential, and $\langle\cdot\rangle_F$ is the normalized Fermi-surface average. The trace identity follows from $\mathrm{Tr}(\Delta'\Delta'^\dagger)=2|q|^2$ and $\det(\Delta'\Delta'^\dagger)=|O|^2$ on each pocket. A Hubbard-Stratonovich decoupling of the term $-b|O|^2$ replaces it by $b|Q|^2-2b\,\mathrm{Re}(Q^*O)$. Completing the square fixes the normalization $\langle Q\rangle=\langle O\rangle$ and gives $r_Q=b$ and $g=-2b$ at the matching scale. Rescaling $Q$ changes both coefficients and preserves $g^2/(4r_Q)=b$.}

{In this matching the residual pair quartic is $2b|q|^4$ and the bare $u$ is zero at the order retained. With the isotropic stiffness the Gaussian zero is $8b k_BT\Pi(0)=1$. In the three-dimensional continuum limit it occurs at $\sqrt{\alpha}=b k_BT/(\pi\kappa_1^{3/2})$. The residual pair interaction produces corrections controlled by the same dimensionless combination $b k_BT\Pi(0)$. In addition, for the auxiliary field introduced here there is no bare $|Q|^4$ term at this matching order. For $|g||Q|<\alpha$, integrating the quadratic pair modes gives $2k_BT\int d^d k/(2\pi)^d\,\ln[1-g^2|Q|^2G_q(\bm k)^2]$. Its $|Q|^4$ coefficient is $-g^4 k_BT\int d^d k\,G_q(\bm k)^4/(2\pi)^d<0$. This weak-coupling estimate yields a local Gaussian mass zero but supplies neither a stabilizing quartic nor a parametric separation from interacting pair fluctuations. A stable pure-quartet interval therefore requires further microscopic input.}

\subsection{Microscopic origin of the second Josephson harmonic
\texorpdfstring{$J_2$}{J2} from coherent \texorpdfstring{$4e$}{4e} tunneling}

Consider two superconductors, labelled \(L\) and \(R\), with superconducting
phases \(\phi_L\) and \(\phi_R\), coupled by a weak tunnel contact:
\begin{subequations}\label{eqS:tunnel-H}
\begin{align}
H &= H_L+H_R+H_T,\\
H_T
&=
\sum_{kp\sigma}
\left(
t_{kp}\,c^\dagger_{Lk\sigma}c_{Rp\sigma}
+
t_{kp}^{\ast}\,c^\dagger_{Rp\sigma}c_{Lk\sigma}
\right).
\label{eqS:HT}
\end{align}
\end{subequations}
Here \(t_{kp}\) is the single-electron tunneling matrix element between a
state \(k\) in the left electrode and a state \(p\) in the right electrode.
We reserve \(\mathcal T\) for time reversal and \(T\) for temperature; the
tunneling operator in Nambu space will be denoted by \(\hat T\).

For quadratic superconducting leads, introduce their full spin-Nambu Green functions and the single-electron tunneling matrix $\hat T$. The combined inverse Green function is
\begin{equation}
\mathcal G^{-1}=\begin{pmatrix}
G_L^{-1}&-\hat T\\-\hat T^\dagger&G_R^{-1}
\end{pmatrix}.
\label{eqS:Ginv-block}
\end{equation}
Using a fully redundant Nambu representation, the fermionic Gaussian integral gives a phase-dependent action $-\tfrac12\mathrm{Tr}\ln\mathcal G^{-1}$, up to phase-independent constants. The factor $1/2$ removes Nambu double counting; a nonredundant reduced representation must use a correspondingly reduced trace. The trace includes spin, Nambu, momentum, Matsubara frequency, and lead indices.

Subtracting the uncoupled-lead contribution and applying the block determinant identity yields
\begin{equation}
S_{\rm eff}=-\frac12\mathrm{Tr}\ln(1-G_L\hat T G_R\hat T^\dagger).
\label{eqS:Seff-log}
\end{equation}
The trace in this equation is over one lead's spin-Nambu and spectral variables. Its weak-tunneling expansion is
\begin{equation}
S_{\rm eff}=\frac12\sum_{n=1}^{\infty}\frac1n
\mathrm{Tr}[(G_L\hat T G_R\hat T^\dagger)^n].
\label{eqS:Seff-expand}
\end{equation}
The junction free energy is $F_J=k_BT S_{\rm eff}$ after the Matsubara trace. This construction fixes the perturbative orders; actual harmonic coefficients still require the lead Green functions and tunnel matrix elements.

\rev{Each factor $G_L\hat T G_R\hat T^\dagger$ contains two
single-electron tunneling events.
For an even phase-dependent junction free energy,
$F_J(\phi)=F_J(-\phi)$, the phase-dependent part of the
$n=1$ term is second order in the tunneling amplitude and
has the form}
{
\begin{equation}
\left.F_J^{(1)}\right|_{\cos\phi}
=-J_1\cos\phi,
\qquad \phi=\phi_L-\phi_R.
\label{eqS:J1-cos}
\end{equation}
}
\rev{Here we use the same gauge convention as in
Sec.~\ref{sec:4e}, with zero vector-potential line integral
across the contact.}
This is the microscopic origin of the Ambegaokar-Baratoff first harmonic.
\rev{At order $|t_{kp}|^4$, the $n=2$ trace contains a phase-independent term, a correction to the first harmonic, and a second harmonic from coherent transfer of two Cooper pairs. The second-harmonic contribution at this order is}
\begin{equation}
\rev{\left.F_J^{(2)}\right|_{\cos(2\phi)}}
=
-J_2\cos(2\phi).
\label{eqS:J2-cos}
\end{equation}
Thus the junction free energy contains
\begin{equation}
F_J(\phi)
=
-J_1\cos\phi
-
J_2\cos(2\phi)
+\cdots .
\label{eqS:FJ-harmonics}
\end{equation}
\rev{The ordinary tunnel expansion already generates a second harmonic.
With a nonzero leading $J_1$, its pair-transfer coefficient $J_2$
is parametrically smaller in the tunnel limit.}
\cite{Golubov2004RMP}

In the tunnel limit it is useful to introduce the dimensionless normal-state
tunneling-strength parameter
\begin{equation}
\mathcal D
\equiv
\pi^2 N_L(0)\,N_R(0)\,|t_T|^2
\ll 1 .
\label{eqS:Ddef}
\end{equation}
\rev{Here $N_L(0)$ and $N_R(0)$ are the normal-state densities
of states per spin and per unit energy at the Fermi level,
in the contact-state normalization used for the hopping.
The energy $t_T$ sets the scale of the tunneling matrix elements:
$t_{kp}=t_Tu_{kp}$ with fixed dimensionless $u_{kp}$.
Thus $\mathcal D$ is dimensionless and scales linearly with
weak channel transparency.
Its conversion to a transmission probability depends on the contact.}

Since \(\mathcal D\propto |t_T|^2\), the leading Josephson coupling scales as
\begin{equation}
J_1\propto |t_T|^2\propto \mathcal D,
\label{eqS:J1-scaling}
\end{equation}
whereas the coherent two-Cooper-pair process scales as
\begin{equation}
J_2\propto |t_T|^4\propto \mathcal D^2.
\label{eqS:J2-scaling}
\end{equation}
\rev{At fixed temperature, when all channel transparencies are
scaled to zero together and the first harmonic has a nonzero
leading coefficient, the ordinary tunnel expansion gives}
\begin{equation}
\frac{J_2}{J_1}
=
\mathcal O(\mathcal D).
\label{eqS:J2J1_scaling}
\end{equation}
\rev{Suppression of the leading $J_1$, for instance near a
$0$-$\pi$ cancellation, or an additional direct quartet-transfer
term can make the two harmonics comparable.}

\rev{For identical isotropic BCS electrodes in the tunnel limit, Ambegaokar and Baratoff obtained the leading critical current~\cite{AmbegaokarBaratoff1963,AmbegaokarBaratoffErratum1963},}
\begin{subequations}\label{eqS:J1J2_explicit}
\begin{align}
I_1(T)
&=
\frac{\pi\Delta(T)}{2eR_N}
\tanh\!\left[\frac{\Delta(T)}{2k_B T}\right],
\label{eqS:I1AB_again}\\
J_1(T)
&=
\frac{\hbar}{2e}\,I_1(T),
\qquad
\rev{\frac{J_2(T)}{J_1(T)}=\mathcal O(\mathcal D).}
\label{eqS:J2_scaling}
\end{align}
\end{subequations}
Here \(T\) is the temperature, \(k_B\) is Boltzmann's constant,
\(\Delta(T)\) is the superconducting gap, and \(R_N\) is the normal-state
junction resistance.

The current-phase relation follows from
\[
I(\phi)
=
\frac{2e}{\hbar}\frac{\partial F_J}{\partial \phi}.
\]
Using Eq.~\eqref{eqS:FJ-harmonics}, this gives
\begin{equation}
I(\phi)
=
I_1\sin\phi
+
I_2\sin(2\phi)
+\cdots,
\qquad
I_n
=
\frac{2e}{\hbar}\,n\,J_n .
\label{eqS:CPR}
\end{equation}
This tunnel expansion concerns the harmonic content of the junction free
energy. It does not rely on the unitary perfect-square form of the bulk
mixed singlet-triplet spectrum discussed in the main text.

\section{\rev{Gap closings, bulk indices and edge states of the Rashba model}}
\label{secS:edge}

For the real pairing amplitudes and time-reversal-invariant normal Hamiltonian in Sec.~\ref{sec:numerics}, the Kramers-Nambu Hamiltonian is
\begin{equation}
\begin{aligned}
H'_s(\bm k)={}&\tau_z\otimes[\xi\Id+\bm g\cdot\bm\sigma]\\
&+\tau_x\otimes[\Delta_s\Id+s\bm d\cdot\bm\sigma],
\end{aligned}
\label{eqS:homotopy-H}
\end{equation}
where $0\le s\le1$, $\bm g=\alpha_R(\sin k_y,-\sin k_x,0)$, $\bm d=\Delta_t(\sin k_x,\sin k_y,0)$, and $\xi=-2t(\cos k_x+\cos k_y)-\mu$. The endpoint $s=1$ is the stated model and $s=0$ has purely singlet pairing. This interpolation preserves time reversal and intrinsic PHS.

In a chiral basis, one off-diagonal block can be chosen as
\begin{equation}
D_s=\xi\Id+\bm g\cdot\bm\sigma
+\ii(\Delta_s\Id+s\bm d\cdot\bm\sigma).
\label{eqS:homotopy-D}
\end{equation}
Because $\bm g\cdot\bm d=0$,
\begin{align}
\det D_s={}&\xi^2-\Delta_s^2
-(\alpha_R^2-s^2\Delta_t^2)(\sin^2k_x+\sin^2k_y)
\nonumber\\
&+2\ii\xi\Delta_s.
\label{eqS:homotopy-det}
\end{align}
A zero-energy bulk state requires $\det D_s=0$. Since $\Delta_s=0.25\ne0$, its imaginary part requires $\xi=0$. At that point the real part is strictly negative for all $s\in[0,1]$, because $\alpha_R=0.6>\Delta_t=0.35$. Hence the gap cannot close along the interpolation. At $s=0$, the gap is at least $|\Delta_s|$ and the normal-state terms may be continuously removed, leaving an atomic singlet BdG Hamiltonian. The \rev{$\Delta_t/t=0.35$ parameter set} is therefore topologically trivial, $\nu=0$.

{
For the full positive-$\Delta_t$ scan, set $s=1$ in Eq.~\eqref{eqS:homotopy-det} and write $R=\sin^2k_x+\sin^2k_y$. The pairing determinant is $\mathrm{Sc}(q^2)=\Delta_s^2-\Delta_t^2R$. Since the two spin vectors are perpendicular, the chiral determinant reads
\begin{equation}
 \det D_1=\xi^2-\alpha_R^2R-\mathrm{Sc}(q^2)+2\ii\xi\Delta_s .
 \label{eqS:rashba-scalar}
\end{equation}
A node at $\Delta_s\ne0$ thus requires both $\xi=0$ and $(\Delta_t^2-\alpha_R^2)R=\Delta_s^2$. In particular the physical gap determinant $\mathrm{Sc}(q^2)=-\alpha_R^2R$ at such a node can be nonzero. The normal-state spin splitting enters the bulk gap-closing condition along with the pairing scalar.

At $\mu=-2t$, the contour $\xi=0$ satisfies $u+v=1$ with $u=\cos k_x$ and $v=\cos k_y$ in $[0,1]$. Along it $R=1+2u(1-u)$ takes every value from 1 to $3/2$. The exact nodal interval is therefore
\begin{equation}
 \sqrt{\alpha_R^2+\tfrac23\Delta_s^2}\ \leq\ \Delta_t\
       \leq\ \sqrt{\alpha_R^2+\Delta_s^2}.
 \label{eqS:rashba-nodal}
\end{equation}
For the values used here this gives $\sqrt{241/600}\leq\Delta_t/t\leq0.65$, with $\sqrt{241/600}\simeq0.633772$. Nodes first occur on the diagonals at $(\pm\pi/3,\pm\pi/3)$ and last occur on the axes at $(\pm\pi/2,0)$ and $(0,\pm\pi/2)$. Below the lower endpoint, decreasing $\Delta_t$ to zero gives the trivial singlet limit.

Above the upper endpoint, keep $\Delta_t$ fixed and scale $(\alpha_R,\Delta_s)\mapsto(\eta\alpha_R,\eta\Delta_s)$, $1\geq\eta\geq0$. For $\eta>0$ a node would require $\xi=0$, where $R\geq1$ and $(\Delta_t^2-\eta^2\alpha_R^2)R-\eta^2\Delta_s^2>0$. At $\eta=0$ the squared excitation energy is $\xi^2+\Delta_t^2R$ and is positive everywhere. The gapped endpoint commutes with $M=\tau_z\otimes\sigma_z$. In its $M=\pm1$ sectors the Hamiltonians are
\begin{equation}
 h_\pm=\Delta_t\sin k_x\,\rho_x
       \ \pm\Delta_t\sin k_y\,\rho_y+\xi\,\rho_z ,
 \label{eqS:rashba-helical}
\end{equation}
where $\rho_i$ are Pauli matrices within each sector. At $(0,0),(\pi,0),(0,\pi),(\pi,\pi)$ the four masses are $(-2,2,2,6)t$ and the corresponding Dirac orientations of $h_+$ are $(+,-,-,+)$. The half-sum of the products of orientation and mass sign has magnitude one. Time reversal exchanges the sectors, so the occupied blocks have opposite unit Chern numbers. The parity of either Chern number gives $\nu=1$. The gap-preserving path carries this index back to finite Rashba coupling and finite singlet amplitude.\cite{QiHughesZhang2009PRL,DeNittisGomi2022DIII}

The gap of the $\Delta_t/t=0.80$ example also admits a simple bound. Split the BdG Hamiltonian into the kinetic plus triplet terms and the Rashba plus singlet terms. Within each part the two terms anticommute. The singular-value triangle inequality then gives
\begin{equation}
 E_{\rm gap}(\bm k)\geq
 \sqrt{\xi^2+\Delta_t^2R}-\sqrt{\alpha_R^2R+\Delta_s^2}.
 \label{eqS:rashba-bound}
\end{equation}
For $R\geq1$ the right-hand side is at least $(0.80-0.65)t\sqrt R$. For $R\leq1$, $(1-\cos k_x)(1-\cos k_y)\geq0$ implies $\xi^2\geq4t^2(1-R)$. The first root is then at least $0.80t$ and the second at most $0.65t$. At $(\pi/2,0)$ the eigenvalue magnitudes are $(0.80\pm0.65)t$. The lower bound is attained and the full-zone gap is $0.15t$.}

\begin{figure}[!tbp]
 \centering
 \includegraphics[width=\linewidth]{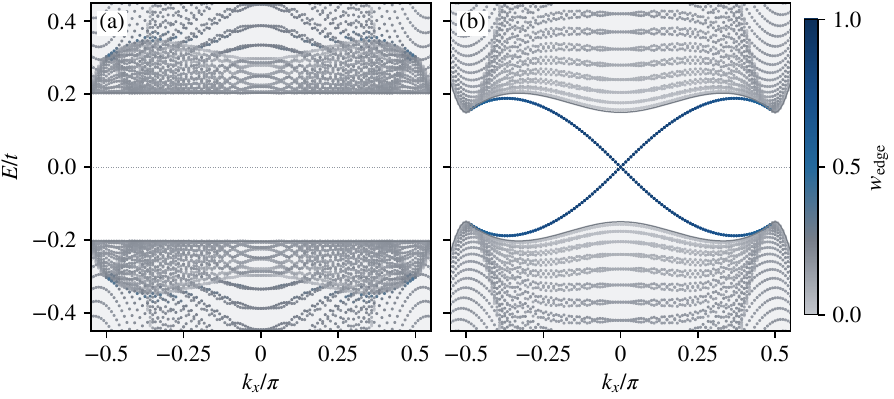}
 \caption{\rev{Edge spectra of strips with $N_y=80$, open along $y$, for (a) $\Delta_t/t=0.35$ and (b) $0.80$. Colour gives the probability in the outermost six rows at each boundary, summed over both edges. The thin grey curves mark the projected bulk gap. The edge-localized branches in (b) meet at $k_x=0$ in the infinite-width limit.}}
 \label{fig:rashba-edge}
\end{figure}

\rev{For a strip of $N_y$ rows, periodic along $x$ and open
along $y$, let $h_y(k_x)$ and $\Delta_y(k_x)$ be the
$2N_y\times2N_y$ normal and pairing matrices on row and spin indices.
In the conventional Nambu basis the strip Hamiltonian is}
\begin{equation}
H_{\rm strip}(k_x)=\begin{pmatrix}
h_y(k_x)&\Delta_y(k_x)\\
\Delta_y^\dagger(k_x)&-h_y^{\mathsf T}(-k_x)
\end{pmatrix}.
\label{eqS:strip-correct}
\end{equation}
{
For the numerical calculation we use the unitarily equivalent
Kramers-Nambu basis of Eq.~\eqref{eq:kramers-nambu},
rotating the hole spinor on each row by $\ii\sigma_y$.
We Fourier transform Eq.~\eqref{eq:BdG-rashba-full}
in $k_y$ and retain the nearest-neighbour bonds within the $N_y$ rows. Closing the strip periodically recovers the bulk Bloch spectrum at the allowed transverse momenta. Figure~\ref{fig:rashba-edge} compares open boundaries at the two triplet amplitudes. At $k_x=0$ in the $\nu=1$ phase, the smallest $|E|/t$ falls from $8.36\times10^{-4}$ at $N_y=40$ to $2.88\times10^{-6}$ at 80 and $9.93\times10^{-9}$ at 120. At the largest width, the four states closest to zero have a mean edge weight of $87.1\%$, summed over the six outermost rows at either boundary. Projecting transverse position into these four states resolves two states on each edge. On each edge the projected $\partial H/\partial k_x$ has eigenvalues $\pm0.26769t$ at $N_y=120$, giving the opposite slopes in the wide-strip limit ($a=1$). In the trivial phase the corresponding energy is $0.20382t$ and the edge weight is $10.5\%$. The width dependence identifies edge hybridization as the origin of the small finite-strip splitting in the nontrivial case.}

\begin{figure}[!htbp]
  \centering
  \includegraphics[width=\linewidth]{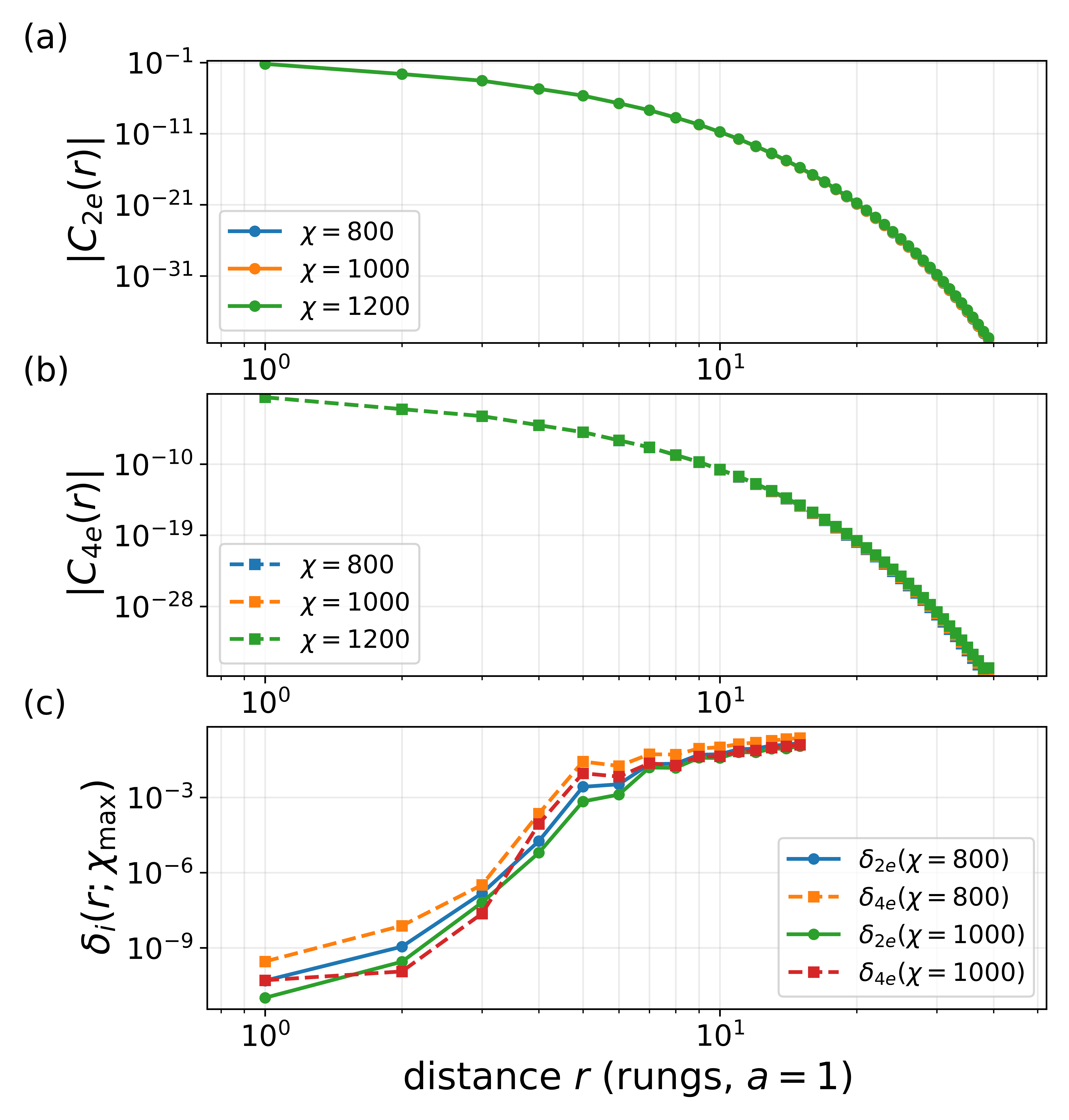}
  \caption{Bond-dimension dependence of the $L=80$, $U=4$ correlators with open boundaries and a central reference rung. Panels (a) and (b) show $|C_{2e}(r)|$ and $|C_{4e}(r)|$ for $\chi_{\max}=800,1000,1200$. Panel (c) shows relative deviations from $\chi_{\rm ref}=1200$ only where the reference magnitude is at least $10^{-16}$. This operational cutoff does not by itself certify accuracy. Table~\ref{tab:chi-deviations} quantifies residual differences above the cutoff, and Eq.~\eqref{eqS:common-tail} compares both channels on an identical distance interval.}
  \label{figS:chi_convergence}
\end{figure}

\section{Two-orbital-chain correlators and bond-dimension convergence}
\label{secS:quartet_chain}

We consider the two-orbital spinful chain used in the main text, with open
boundary conditions, half filling, and identical Hamiltonian parameters. Ground
states and correlators are computed with DMRG in its matrix-product-state (MPS)
formulation.\cite{White1992,White1993,Schollwock2011,HauschildPollmann2018TenPy}

In the implementation, the four fermionic modes of each rung are ordered as separate MPS sites, so the MPS has $M=4L$ sites with local occupation $s_\ell\in\{0,1\}$. It writes the many-body state as
\begin{equation}
\ket{\psi}
=
\sum_{\{s_i\}}
A^{[1]s_1}A^{[2]s_2}\cdots A^{[M]s_M}
\ket{s_1s_2\cdots s_M},
\label{eqS:MPS_def}
\end{equation}
where each \(A^{[i]s_i}\) is a matrix of size
\(\chi_{i-1}\times\chi_i\). The integers \(\chi_i\) are the bond dimensions:
they control how much bipartite entanglement can be represented across the cut
between sites \(i\) and \(i+1\).\cite{Schollwock2011} Equivalently, across that
cut the Schmidt decomposition reads
\begin{equation}
\ket{\psi}
=
\sum_{\alpha=1}^{\chi_i}
\lambda^{(i)}_\alpha
\ket{\alpha_L^{(i)}}\ket{\alpha_R^{(i)}},
\label{eqS:Schmidt_def}
\end{equation}
so \(\chi_i\) is the number of kept Schmidt states. In practical DMRG
calculations one enforces \(\chi_i\le \chi_{\max}\) for all cuts. Increasing $\chi_{\max}$ enlarges the variational space, but convergence of each correlator must be checked explicitly.\cite{White1992,White1993,Schollwock2011}

\rev{For the operators in Eqs.~\eqref{eq:Delta_i_1d} and
\eqref{eq:Q_i_1d}, fermionic anticommutation gives}
\begin{equation}
\Delta_i^2=-2Q_i,
\label{eqS:Delta_square_Q}
\end{equation}
so \(Q_i\) is, up to a fixed numerical factor, the square of the local pair
field. This realizes a local charge-$4e$ composite of the chosen pair operator. It has the same gauge charge as the selected GL scalar channel; the operator normalization remains explicit.

\rev{We evaluate the equal-time correlators defined in
Eq.~\eqref{eq:C2e_C4e_def}.}
\rev{For the central-reference results we take $i=L/2$ and $1\le r\le L/2-1$. The origin-averaged calculation below uses all $L-r$ pairs, including those touching an end.}

\rev{We compare the magnitudes at each separation to track their dependence on bond dimension,}
\begin{equation}
\widetilde C_i(r;\chi_{\max})
\equiv
\big|C_i(r;\chi_{\max})\big|,
\qquad
i\in\{2e,4e\}.
\label{eqS:Cabs}
\end{equation}
We perform a bond-dimension scan at fixed \((U,L)=(4.0,80)\) using
\begin{equation}
\chi_{\max}=800,\ 1000,\ 1200,
\label{eqS:chi_scan}
\end{equation}
and take the largest value,
\begin{equation}
\chi_{\rm ref}=1200,
\label{eqS:chi_ref}
\end{equation}
as the reference calculation.

The pointwise relative deviation from the reference is defined as
\begin{equation}
\delta_i(r;\chi_{\max})
=
\frac{
\big|
\widetilde C_i(r;\chi_{\max})
-
\widetilde C_i(r;\chi_{\rm ref})
\big|
}{
\widetilde C_i(r;\chi_{\rm ref})
}.
\label{eqS:reldev}
\end{equation}
For $i\in\{2e,4e\}$, we report relative deviations only where the reference magnitude exceeds the operational cutoff
\begin{equation}
\widetilde C_i(r;\chi_{\rm ref})\ge C_{\rm floor},
\qquad C_{\rm floor}=10^{-16}.
\label{eqS:floor}
\end{equation}
The cutoff restricts the comparison to a chosen amplitude range; it is not a universal noise floor set by double precision. A small correlator may be represented with relative accuracy, and convergence cannot be inferred from its absolute magnitude or from nearly overlapping logarithmic curves.

For the stored $L=80$, $U=4$ calculations, both channels exceed $10^{-16}$ through $r=15$ at all three bond dimensions. The pointwise deviations between $\chi=800$ and the $\chi=1200$ reference are summarized in Table~\ref{tab:chi-deviations}.

\begin{table}[h]
\centering
\caption{Relative correlator differences $100\delta_i(r;800)$ from the $\chi_{\rm ref}=1200$ data. Both correlators exceed the adopted magnitude cutoff at every listed separation.}
\label{tab:chi-deviations}
\begin{tabular}{ccc}
\toprule
$r$ & $2e$ channel (\%) & $4e$ channel (\%)\\
\midrule
5 & 0.26 & 2.68\\
10 & 5.24 & 10.00\\
15 & 15.05 & 23.14\\
\bottomrule
\end{tabular}
\end{table}

\rev{The residual changes grow overall with separation and are
already measurable above the cutoff.} The energies are much less sensitive to bond dimension, so energy convergence alone does not certify these small correlators.

As a common-window check, define the tail sums of Eq.~\eqref{eq:Stail_def_main} using only the identical interval $10\le r\le15$ in both channels and all three calculations. The ratios are
\begin{equation}
R_{\rm tail}^{[10,15]}=
\begin{cases}
1.157,&\chi_{\max}=800,\\
1.211,&\chi_{\max}=1000,\\
1.220,&\chi_{\max}=1200.
\end{cases}
\label{eqS:common-tail}
\end{equation}
\rev{For these operator normalizations the common-window ratio exceeds unity at all three bond dimensions. Figure~\ref{figS:chi_convergence} also shows the raw magnitudes beyond this window.}

\paragraph{\rev{Dependence on chain length and choice of reference rung.}}
\rev{At fixed $\chi=800$ and common centre $r=5$, $\eta_{2e}-\eta_{4e}$ is 0.50 to two decimal places for each of $L=24,32,40,60,80$. At $r=10$, where a seven-point window is available for $L=32,40,60,80$, the differences are $2.00,1.91,1.93,1.93$, respectively. The $L=16$ central-reference data stop at $r=7$, short of the first full window. These comparisons test length dependence at fixed separation.}

\rev{For independent origin averages, we compute the signed correlator for every pair of rungs in an open chain before taking the magnitude,}
{
\begin{equation}
\overline C_O(r)=\frac{1}{L-r}\sum_{i=0}^{L-r-1}
   \langle O_i^\dagger O_{i+r}\rangle ,\qquad O=\Delta,Q .
\label{eqS:origin-average}
\end{equation}}
\rev{The DMRG conserves particle number and starts from alternating full and empty rungs. We group the four modes on each rung for the measurement, including the internal fermionic signs. Table~\ref{tab:averaged-slopes} gives the slopes at the two end centres of the comparison. At $L=24$ and $\chi=128$ the pair slope is smaller at all five resolved centres ($r=5$-9). At $\chi=256$ the quartet slope is smaller throughout those same centres. At $L=24$, $\chi=480$, and $L=32$, $\chi=384$, the quartet slope is smaller at every centre from 5 through 10. If the origin-averaged pair slope stays equal to or below the quartet slope at larger $L$ and $\chi$ over a resolved interval, the quartet-decay advantage is absent on that interval.}

\begin{table}[!htb]
\centering
\caption{\rev{Local logarithmic slopes of the rung-averaged correlators from Eq.~\eqref{eqS:origin-average}. Each fit uses seven separations. A dash indicates that at least one magnitude in that channel's window falls below $10^{-14}$.}}
\label{tab:averaged-slopes}
{
\begin{tabular}{rr cc cc}
\toprule
 & & \multicolumn{2}{c}{$r=5$} & \multicolumn{2}{c}{$r=10$}\\
\cmidrule(lr){3-4}\cmidrule(lr){5-6}
$L$ & $\chi_{\max}$ & $\eta_{2e}$ & $\eta_{4e}$ & $\eta_{2e}$ & $\eta_{4e}$\\
\midrule
16 & 192 & 10.36 & 9.94 & \textemdash & 21.44\\
24 & 128 & 10.64 & 10.72 & \textemdash & \textemdash\\
24 & 256 & 10.31 & 9.94 & \textemdash & 21.82\\
24 & 480 & 10.18 & 9.46 & 22.75 & 20.55\\
32 & 256 & 10.32 & 10.02 & \textemdash & 21.99\\
32 & 384 & 10.23 & 9.61 & 22.97 & 21.17\\
\bottomrule
\end{tabular}}
\end{table}

\paragraph{\rev{Comparison with published infinite-chain data.}}
\rev{We apply Eq.~\eqref{eq:eta_fit_main} to the bare, single-component interorbital pair and local quartet correlators published for Fig.~A7 of Ref.~\cite{Soldini2024PRB} (source data, DOI \href{https://doi.org/10.6084/m9.figshare.25663788}{10.6084/m9.figshare.25663788}). At their quartet point $(\delta,t,V/U)=(-0.5,0.6,0.2)$, in units $-\mu=1$, the $\chi=700$ slopes $(\eta_{2e},\eta_{4e})$ are $(3.85,1.41)$ and $(7.90,1.54)$ at centres $5$ and $10$. At their charge-density-wave point $(0,1,0.2)$ the corresponding values are $(1.59,2.30)$ and $(3.00,3.96)$. Here $U$ and $V$ denote their intraorbital and interorbital attractions and $\delta=2+U+2V$. The weak variation of the quartet slope at their quartet point differs markedly from its rise in Fig.~\ref{fig:eta-rtail-1d}(a). Since these are different Hamiltonians, ensembles and pair operators, the benchmark compares the decay diagnostic across states. For our open chain the resolved interval establishes a difference between the two slopes. Longer-distance data are needed to identify an asymptotic decay law.}

\FloatBarrier

\bibliographystyle{apsrev4-2}
\bibliography{main}

@article{GorkovRashba2001,
  author        = {Gor'kov, L. P. and Rashba, E. I.},
  title         = {Superconducting 2D System with Lifted Spin Degeneracy: Mixed Singlet-Triplet State},
  journal       = {Physical Review Letters},
  volume        = {87},
  number        = {3},
  pages         = {037004},
  year          = {2001},
  doi           = {10.1103/PhysRevLett.87.037004},
  url           = {https://doi.org/10.1103/PhysRevLett.87.037004}
}

@article{Frigeri2004,
  author        = {Frigeri, P. A. and Agterberg, D. F. and Koga, A. and Sigrist, M.},
  title         = {Superconductivity without Inversion Symmetry: {MnSi} versus {CePt$_3$Si}},
  journal       = {Physical Review Letters},
  volume        = {92},
  number        = {9},
  pages         = {097001},
  year          = {2004},
  doi           = {10.1103/PhysRevLett.92.097001},
  url           = {https://doi.org/10.1103/PhysRevLett.92.097001}
}

@article{Smidman2017,
  author        = {Smidman, M. and Salamon, M. B. and Yuan, H. Q. and Agterberg, D. F.},
  title         = {Superconductivity and Spin--Orbit Coupling in Non-Centrosymmetric Materials: A Review},
  journal       = {Reports on Progress in Physics},
  volume        = {80},
  number        = {3},
  pages         = {036501},
  year          = {2017},
  doi           = {10.1088/1361-6633/80/3/036501},
  url           = {https://doi.org/10.1088/1361-6633/80/3/036501}
}

@article{Fischer2023,
  author        = {Fischer, Mark H. and Sigrist, Manfred and Agterberg, Daniel F. and Yanase, Youichi},
  title         = {Superconductivity and Local Inversion-Symmetry Breaking},
  journal       = {Annual Review of Condensed Matter Physics},
  volume        = {14},
  pages         = {153--172},
  year          = {2023},
  doi           = {10.1146/annurev-conmatphys-040521-042511},
  url           = {https://doi.org/10.1146/annurev-conmatphys-040521-042511}
}

@article{Nambu1960,
  author        = {Nambu, Yoichiro},
  title         = {Quasi-Particles and Gauge Invariance in the Theory of Superconductivity},
  journal       = {Physical Review},
  volume        = {117},
  number        = {3},
  pages         = {648--663},
  year          = {1960},
  doi           = {10.1103/PhysRev.117.648},
  url           = {https://doi.org/10.1103/PhysRev.117.648}
}

@book{deGennes1966,
  author        = {de Gennes, Pierre-Gilles},
  title         = {Superconductivity of Metals and Alloys},
  publisher     = {W. A. Benjamin},
  address       = {New York},
  year          = {1966},
  url           = {https://openlibrary.org/books/OL5945521M/Superconductivity_of_metals_and_alloys}
}

@article{AltlandZirnbauer1997,
  author        = {Altland, Alexander and Zirnbauer, Martin R.},
  title         = {Nonstandard symmetry classes in mesoscopic normal-superconducting hybrid structures},
  journal       = {Physical Review B},
  volume        = {55},
  number        = {2},
  pages         = {1142--1161},
  year          = {1997},
  doi           = {10.1103/PhysRevB.55.1142},
  url           = {https://doi.org/10.1103/PhysRevB.55.1142}
}

@article{SchnyderRyuFurusakiLudwig2008,
  author        = {Schnyder, Andreas P. and Ryu, Shinsei and Furusaki, Akira and Ludwig, Andreas W. W.},
  title         = {Classification of topological insulators and superconductors in three spatial dimensions},
  journal       = {Physical Review B},
  volume        = {78},
  number        = {19},
  pages         = {195125},
  year          = {2008},
  doi           = {10.1103/PhysRevB.78.195125},
  url           = {https://doi.org/10.1103/PhysRevB.78.195125}
}

@article{Hatsugai2010,
  author        = {Hatsugai, Yasuhiro},
  title         = {Symmetry-protected $\mathbb{Z}_2$-quantization and quaternionic Berry connection with Kramers degeneracy},
  journal       = {New Journal of Physics},
  volume        = {12},
  pages         = {065004},
  year          = {2010},
  doi           = {10.1088/1367-2630/12/6/065004},
  url           = {https://doi.org/10.1088/1367-2630/12/6/065004}
}

@article{DeNittisGomi2015CMP,
  author        = {De Nittis, Giuseppe and Gomi, Kiyonori},
  title         = {Classification of ``Quaternionic'' Bloch-bundles: Topological Quantum Systems of type AII},
  journal       = {Communications in Mathematical Physics},
  volume        = {339},
  number        = {1},
  pages         = {1--55},
  year          = {2015},
  doi           = {10.1007/s00220-015-2390-0},
  url           = {https://doi.org/10.1007/s00220-015-2390-0}
}

@article{BergFradkinKivelson2009NatPhys,
  author        = {Berg, Erez and Fradkin, Eduardo and Kivelson, Steven A.},
  title         = {Charge-4e superconductivity from pair-density-wave order in certain high-temperature superconductors},
  journal       = {Nature Physics},
  volume        = {5},
  number        = {11},
  pages         = {830--833},
  year          = {2009},
  doi           = {10.1038/nphys1389},
  url           = {https://doi.org/10.1038/nphys1389}
}

@article{Agterberg2020ARCM,
  author        = {Agterberg, Daniel F. and Davis, J. C. S{\'e}amus and Edkins, Stephen D. and Fradkin, Eduardo and Van Harlingen, Dale J. and Kivelson, Steven A. and Lee, Patrick A. and Radzihovsky, Leo and Tranquada, John M. and Wang, Yuxuan},
  title         = {The Physics of Pair-Density Waves: Cuprate Superconductors and Beyond},
  journal       = {Annual Review of Condensed Matter Physics},
  volume        = {11},
  pages         = {231--270},
  year          = {2020},
  doi           = {10.1146/annurev-conmatphys-031119-050711},
  url           = {https://doi.org/10.1146/annurev-conmatphys-031119-050711}
}

@article{WuWang2024npjQM,
  author        = {Wu, Yi-Ming and Wang, Yuxuan},
  title         = {$d$-wave charge-4e superconductivity from fluctuating pair density waves},
  journal       = {npj Quantum Materials},
  volume        = {9},
  pages         = {66},
  year          = {2024},
  doi           = {10.1038/s41535-024-00674-y},
  url           = {https://doi.org/10.1038/s41535-024-00674-y}
}

@article{Soldini2024PRB,
  author        = {Soldini, Martina O. and Fischer, Mark H. and Neupert, Titus},
  title         = {Charge-4e superconductivity in a Hubbard model},
  journal       = {Physical Review B},
  volume        = {109},
  pages         = {214509},
  year          = {2024},
  doi           = {10.1103/PhysRevB.109.214509},
  url           = {https://doi.org/10.1103/PhysRevB.109.214509}
}

@article{PhysRevB.82.134511,
  author        = {Herland, Egil V. and Babaev, Egor and Sudb\o{}, Asle},
  title         = {Phase transitions in a three dimensional $U(1)\ifmmode\times\else\texttimes\fi{}U(1)$ lattice London superconductor: Metallic superfluid and charge-$4e$ superconducting states},
  journal       = {Phys. Rev. B},
  volume        = {82},
  number        = {13},
  pages         = {134511},
  publisher     = {American Physical Society},
  year          = {2010},
  month         = {Oct},
  doi           = {10.1103/PhysRevB.82.134511},
  url           = {https://doi.org/10.1103/PhysRevB.82.134511}
}

@article{Ciaccia2024CommPhys,
  author        = {Ciaccia, Carlo and Haller, Roy and Drachmann, Asbj{\o}rn C. C. and Lindemann, Tyler and Manfra, Michael J. and Schrade, Constantin and Sch{\"o}nenberger, Christian},
  title         = {Charge-4e supercurrent in a two-dimensional InAs--Al superconductor--semiconductor heterostructure},
  journal       = {Communications Physics},
  volume        = {7},
  pages         = {41},
  year          = {2024},
  doi           = {10.1038/s42005-024-01531-x},
  url           = {https://doi.org/10.1038/s42005-024-01531-x}
}

@article{QiHughesZhang2009PRL,
  author        = {Qi, Xiao-Liang and Hughes, Taylor L. and Raghu, S. and Zhang, Shou-Cheng},
  title         = {Time-Reversal-Invariant Topological Superconductors and Superfluids in Two and Three Dimensions},
  journal       = {Physical Review Letters},
  volume        = {102},
  pages         = {187001},
  year          = {2009},
  doi           = {10.1103/PhysRevLett.102.187001},
  url           = {https://doi.org/10.1103/PhysRevLett.102.187001}
}

@article{ZhangKaneMele2013PRL,
  author        = {Zhang, Fan and Kane, Charles L. and Mele, Eugene J.},
  title         = {Time-Reversal-Invariant Topological Superconductivity and Majorana Kramers Pairs},
  journal       = {Physical Review Letters},
  volume        = {111},
  pages         = {056402},
  year          = {2013},
  doi           = {10.1103/PhysRevLett.111.056402},
  url           = {https://doi.org/10.1103/PhysRevLett.111.056402}
}

@article{Liu2017PNAS,
  author        = {Liu, Hanyu and Naumov, Ivan I. and Hoffmann, Roald and Ashcroft, N. W. and Hemley, Russell J.},
  title         = {Potential high-$T_c$ superconducting lanthanum and yttrium hydrides at high pressure},
  journal       = {Proceedings of the National Academy of Sciences},
  volume        = {114},
  number        = {27},
  pages         = {6990--6995},
  year          = {2017},
  doi           = {10.1073/pnas.1704505114},
  url           = {https://doi.org/10.1073/pnas.1704505114}
}

@article{Somayazulu2019PRL,
  author        = {Somayazulu, Maddury and Ahart, Muhtar and Mishra, Ajay K. and Geballe, Zachary M. and Baldini, Maria and Meng, Yue and Struzhkin, Viktor V. and Hemley, Russell J.},
  title         = {Evidence for Superconductivity above 260 K in Lanthanum Superhydride at Megabar Pressures},
  journal       = {Physical Review Letters},
  volume        = {122},
  number        = {2},
  pages         = {027001},
  year          = {2019},
  doi           = {10.1103/PhysRevLett.122.027001},
  url           = {https://doi.org/10.1103/PhysRevLett.122.027001}
}

@article{PhysRevB.102.024501,
  author        = {Elatresh, S. F. and Timusk, T. and Nicol, E. J.},
  title         = {Optical properties of superconducting pressurized ${\mathrm{LaH}}_{10}$},
  journal       = {Phys. Rev. B},
  volume        = {102},
  number        = {2},
  pages         = {024501},
  publisher     = {American Physical Society},
  year          = {2020},
  month         = {Jul},
  doi           = {10.1103/PhysRevB.102.024501},
  url           = {https://doi.org/10.1103/PhysRevB.102.024501}
}

@article{c7w9-7tgy,
  author        = {Tantardini, Christian and Kvashnin, Alexander G. and Giantomassi, Matteo and Ilia\ifmmode \check{s}\else \v{s}\fi{}, Miroslav and Yakobson, Boris I. and Hemley, Russell J. and Gonze, Xavier},
  title         = {Charge density waves and structural phase transition in the high-${T}_{c}$ superconducting ${\mathrm{LaH}}_{10}$ quantum crystal},
  journal       = {Phys. Rev. B},
  volume        = {112},
  number        = {11},
  pages         = {115154},
  publisher     = {American Physical Society},
  year          = {2025},
  month         = {Sep},
  doi           = {10.1103/c7w9-7tgy},
  url           = {https://doi.org/10.1103/c7w9-7tgy}
}

@article{Struzhkin1997Nature,
  author        = {Struzhkin, Viktor V. and Hemley, Russell J. and Mao, Ho-kwang and Timofeev, Yuri A.},
  title         = {Superconductivity at 10--17 K in compressed sulphur},
  journal       = {Nature},
  volume        = {390},
  pages         = {382--384},
  year          = {1997},
  doi           = {10.1038/37074},
  url           = {https://doi.org/10.1038/37074}
}

@article{Boeri2022Roadmap,
  author        = {Boeri, Lilia and Hennig, Richard and Hirschfeld, Peter and Profeta, Gianni and Sanna, Antonio and Zurek, Eva and Pickett, Warren E. and Amsler, Maximilian and Dias, Ranga and Eremets, Mikhail I. and Heil, Christoph and Hemley, Russell J. and Liu, Hanyu and Ma, Yanming and others},
  title         = {The 2021 room-temperature superconductivity roadmap},
  journal       = {Journal of Physics: Condensed Matter},
  volume        = {34},
  number        = {18},
  pages         = {183002},
  year          = {2022},
  doi           = {10.1088/1361-648X/ac2864},
  url           = {https://doi.org/10.1088/1361-648X/ac2864}
}

@article{babaev_2024,
  author        = {Babaev, Egor and Sudb{\o}, Asle and Ashcroft, N. W.},
  title         = {A superconductor to superfluid phase transition in liquid metallic hydrogen},
  journal       = {Nature},
  volume        = {431},
  number        = {7009},
  pages         = {666--668},
  year          = {2004},
  doi           = {10.1038/nature02910},
  url           = {https://doi.org/10.1038/nature02910}
}

@book{SakuraiQM3,
  author        = {J. J. Sakurai and Jim Napolitano},
  title         = {Modern Quantum Mechanics},
  edition       = {3},
  publisher     = {Cambridge University Press},
  address       = {Cambridge},
  year          = {2020},
  doi           = {10.1017/9781108587280},
  url           = {https://doi.org/10.1017/9781108587280},
  isbn          = {9781108473224}
}

@article{SigristUeda1991,
  author        = {Sigrist, Manfred and Ueda, Kazuo},
  title         = {Phenomenological theory of unconventional superconductivity},
  journal       = {Reviews of Modern Physics},
  volume        = {63},
  number        = {2},
  pages         = {239--311},
  year          = {1991},
  doi           = {10.1103/RevModPhys.63.239},
  url           = {https://doi.org/10.1103/RevModPhys.63.239}
}

@article{ChiuTeoSchnyderRyu2016RMP,
  author        = {Ching-Kai Chiu and Jeffrey C. Y. Teo and Andreas P. Schnyder and Shinsei Ryu},
  title         = {Classification of topological quantum matter with symmetries},
  journal       = {Reviews of Modern Physics},
  volume        = {88},
  number        = {3},
  pages         = {035005},
  year          = {2016},
  doi           = {10.1103/RevModPhys.88.035005},
  url           = {https://doi.org/10.1103/RevModPhys.88.035005}
}

@inproceedings{Kitaev2009AIP,
  author        = {Alexei Kitaev},
  title         = {Periodic table for topological insulators and superconductors},
  booktitle     = {Advances in Theoretical Physics: Landau Memorial Conference},
  series        = {AIP Conference Proceedings},
  volume        = {1134},
  pages         = {22--30},
  publisher     = {American Institute of Physics},
  year          = {2009},
  doi           = {10.1063/1.3149495},
  url           = {https://doi.org/10.1063/1.3149495}
}

@article{RyuSchnyderFurusakiLudwig2010NJP,
  author        = {Shinsei Ryu and Andreas P. Schnyder and Akira Furusaki and Andreas W. W. Ludwig},
  title         = {Topological insulators and superconductors: Tenfold way and dimensional hierarchy},
  journal       = {New Journal of Physics},
  volume        = {12},
  pages         = {065010},
  year          = {2010},
  doi           = {10.1088/1367-2630/12/6/065010},
  url           = {https://doi.org/10.1088/1367-2630/12/6/065010}
}

@article{DengWangDuan2014PRB,
  author        = {Deng, Dong-Ling and Wang, Sheng-Tao and Duan, Lu-Ming},
  title         = {Systematic construction of tight-binding Hamiltonians for topological insulators and superconductors},
  journal       = {Physical Review B},
  volume        = {89},
  pages         = {075126},
  year          = {2014},
  doi           = {10.1103/PhysRevB.89.075126},
  url           = {https://doi.org/10.1103/PhysRevB.89.075126}
}

@article{Edelstein1995PRL,
  author        = {Edelstein, Victor M.},
  title         = {Magnetoelectric Effect in Polar Superconductors},
  journal       = {Physical Review Letters},
  volume        = {75},
  number        = {10},
  pages         = {2004--2007},
  year          = {1995},
  month         = {sep},
  doi           = {10.1103/PhysRevLett.75.2004},
  url           = {https://doi.org/10.1103/PhysRevLett.75.2004}
}

@article{KaurAgterbergSigrist2005PRL,
  author        = {R. P. Kaur and D. F. Agterberg and M. Sigrist},
  title         = {Helical Vortex Phase in the Noncentrosymmetric {CePt$_3$Si}},
  journal       = {Physical Review Letters},
  volume        = {94},
  pages         = {137002},
  year          = {2005},
  doi           = {10.1103/PhysRevLett.94.137002},
  url           = {https://doi.org/10.1103/PhysRevLett.94.137002}
}

@book{TinkhamBook,
  author        = {Michael Tinkham},
  title         = {Introduction to Superconductivity},
  edition       = {2},
  publisher     = {McGraw--Hill},
  address       = {New York},
  year          = {1996},
  url           = {https://openlibrary.org/books/OL789944M/Introduction_to_superconductivity},
  isbn          = {9780070648784}
}

@article{FuKane2006TRPolarization,
  author        = {Fu, Liang and Kane, C. L.},
  title         = {Time reversal polarization and a $\mathbb{Z}_2$ adiabatic spin pump},
  journal       = {Physical Review B},
  volume        = {74},
  pages         = {195312},
  year          = {2006},
  doi           = {10.1103/PhysRevB.74.195312},
  url           = {https://doi.org/10.1103/PhysRevB.74.195312}
}

@article{SoluyanovVanderbilt2011,
  author        = {Alexey A. Soluyanov and David Vanderbilt},
  title         = {Computing topological invariants without inversion symmetry},
  journal       = {Phys. Rev. B},
  volume        = {83},
  pages         = {235401},
  year          = {2011},
  doi           = {10.1103/PhysRevB.83.235401},
  url           = {https://doi.org/10.1103/PhysRevB.83.235401}
}

@article{YuQiBernevigFangDaiVanderbilt2011,
  author        = {Yu, Rui and Qi, Xiao Liang and Bernevig, Andrei and Fang, Zhong and Dai, Xi},
  title         = {Equivalent expression of {$\mathbb{Z}_2$} topological invariant for band insulators using the non-Abelian Berry connection},
  journal       = {Phys. Rev. B},
  volume        = {84},
  pages         = {075119},
  year          = {2011},
  doi           = {10.1103/PhysRevB.84.075119},
  url           = {https://doi.org/10.1103/PhysRevB.84.075119}
}

@article{FuKane2007Inversion,
  author        = {Fu, Liang and Kane, C. L.},
  title         = {Topological insulators with inversion symmetry},
  journal       = {Physical Review B},
  volume        = {76},
  pages         = {045302},
  year          = {2007},
  doi           = {10.1103/PhysRevB.76.045302},
  url           = {https://doi.org/10.1103/PhysRevB.76.045302}
}

@article{FuBerg2010OddParity,
  author        = {Liang Fu and Erez Berg},
  title         = {Odd-parity topological superconductors: Theory and application to Cu$_x$Bi$_2$Se$_3$},
  journal       = {Phys. Rev. Lett.},
  volume        = {105},
  pages         = {097001},
  year          = {2010},
  doi           = {10.1103/PhysRevLett.105.097001},
  url           = {https://doi.org/10.1103/PhysRevLett.105.097001}
}

@article{Sato2010OddParity,
  author        = {Sato, Masatoshi},
  title         = {Topological odd-parity superconductors},
  journal       = {Physical Review B},
  volume        = {81},
  number        = {22},
  pages         = {220504{\rev{(R)}}},
  year          = {2010},
  doi           = {10.1103/PhysRevB.81.220504},
  url           = {https://doi.org/10.1103/PhysRevB.81.220504}
}

@article{Schnyder2012PRB,
  author        = {Andreas P. Schnyder and Shinsei Ryu},
  title         = {Topological phases and surface flat bands in superconductors without inversion symmetry},
  journal       = {Phys. Rev. B},
  volume        = {84},
  pages         = {060504(R)},
  year          = {2011},
  doi           = {10.1103/PhysRevB.84.060504},
  url           = {https://doi.org/10.1103/PhysRevB.84.060504}
}

@article{FernandesFu2021PRL,
  author        = {Fernandes, Rafael M. and Fu, Liang},
  title         = {Charge-{$4e$} Superconductivity from Multicomponent Nematic Pairing: Application to Twisted Bilayer Graphene},
  journal       = {Physical Review Letters},
  volume        = {127},
  number        = {4},
  pages         = {047001},
  year          = {2021},
  doi           = {10.1103/PhysRevLett.127.047001},
  url           = {https://doi.org/10.1103/PhysRevLett.127.047001}
}

@article{White1992,
  author        = {White, Steven R.},
  title         = {Density Matrix Formulation for Quantum Renormalization Groups},
  journal       = {Phys. Rev. Lett.},
  volume        = {69},
  pages         = {2863--2866},
  year          = {1992},
  doi           = {10.1103/PhysRevLett.69.2863},
  url           = {https://doi.org/10.1103/PhysRevLett.69.2863}
}

@article{White1993,
  author        = {White, Steven R.},
  title         = {Density-matrix algorithms for quantum renormalization groups},
  journal       = {Phys. Rev. B},
  volume        = {48},
  pages         = {10345--10356},
  year          = {1993},
  doi           = {10.1103/PhysRevB.48.10345},
  url           = {https://doi.org/10.1103/PhysRevB.48.10345}
}

@article{Schollwock2011,
  author        = {Schollw{\"o}ck, Ulrich},
  title         = {The density-matrix renormalization group in the age of matrix product states},
  journal       = {Annals of Physics},
  volume        = {326},
  number        = {1},
  pages         = {96--192},
  year          = {2011},
  doi           = {10.1016/j.aop.2010.09.012},
  url           = {https://doi.org/10.1016/j.aop.2010.09.012}
}

@article{HauschildPollmann2018TenPy,
  author        = {Hauschild, Johannes and Pollmann, Frank},
  title         = {Efficient numerical simulations with tensor networks: Tensor Network Python ({TeNPy})},
  journal       = {SciPost Phys. Lect. Notes},
  volume        = {{\rev{5}}},
  pages         = {},
  year          = {2018},
  doi           = {10.21468/SciPostPhysLectNotes.5},
  url           = {https://doi.org/10.21468/SciPostPhysLectNotes.5},
  eprint        = {1805.00055},
  archiveprefix = {arXiv},
  primaryclass  = {cond-mat.str-el}
}

@article{Golubov2004RMP,
  author        = {Golubov, A. A. and Kupriyanov, M. Yu. and Il'ichev, E.},
  title         = {The current-phase relation in Josephson junctions},
  journal       = {Reviews of Modern Physics},
  volume        = {76},
  number        = {2},
  pages         = {411--469},
  year          = {2004},
  doi           = {10.1103/RevModPhys.76.411},
  url           = {https://doi.org/10.1103/RevModPhys.76.411}
}

@book{ClarkeBraginskiSQUID,
  editor        = {John Clarke and Alex I. Braginski},
  title         = {The SQUID Handbook: Fundamentals and Technology of SQUIDs and SQUID Systems},
  volume        = {1},
  publisher     = {Wiley-VCH},
  address       = {Weinheim},
  year          = {2004},
  doi           = {10.1002/3527603646},
  url           = {https://doi.org/10.1002/3527603646},
  isbn          = {978-3-527-40229-8}
}

@article{AmbegaokarBaratoff1963,
  author = {Ambegaokar, Vinay and Baratoff, Alexis},
  title = {Tunneling Between Superconductors},
  journal = {Phys. Rev. Lett.},
  volume = {10},
  pages = {486--489},
  year = {1963},
  doi = {10.1103/PhysRevLett.10.486}
}

@article{AmbegaokarBaratoffErratum1963,
  author = {Ambegaokar, Vinay and Baratoff, Alexis},
  title = {Erratum: Tunneling Between Superconductors},
  journal = {Phys. Rev. Lett.},
  volume = {11},
  pages = {104},
  year = {1963},
  doi = {10.1103/PhysRevLett.11.104}
}

@article{Cui2026Spin1,
  author = {Cui, Xiaxia and Zhou, Xinchao and Tang, Qiang and Tang, Jau},
  title = {A Phenomenological Effective-Field Theory for a Charged Spin-1 Condensate in Anisotropic Layered Superconductors},
  journal = {Physchem},
  volume = {6},
  pages = {57},
  year = {2026},
  doi = {10.3390/physchem6030057}
}

@article{RantanenEltsov2023,
  author = {Rantanen, R. and Eltsov, V. B.},
  title = {Transition in vortex skyrmion structures in superfluid {$^3$He-A} driven by an analog of the zero-charge effect},
  journal = {Phys. Rev. B},
  volume = {107},
  pages = {104505},
  year = {2023},
  doi = {10.1103/PhysRevB.107.104505}
}

@article{DeNittisGomi2022DIII,
  author = {De Nittis, Giuseppe and Gomi, K.},
  title = {The Cohomology Invariant for Class {DIII} Topological Insulators},
  journal = {Ann. Henri Poincar{\'e}},
  volume = {23},
  pages = {3587--3632},
  year = {2022},
  doi = {10.1007/s00023-022-01183-z}
}

\end{document}